\documentclass[prb,aps,twocolumn,showpacs,superscriptaddress,amsmath,amssymb]{revtex4-1}
\usepackage{graphicx,amsmath,bm,color}
\usepackage{ulem} 
\usepackage{color}
\pdfoutput=1

\newcommand{\beq}{\begin{equation}}
\newcommand{\eeq}{\end{equation}}
\newcommand{\eps}{\epsilon}

\definecolor{grey}{rgb}{.6,.6,.6}
\definecolor{darkgreen}{rgb}{.3,1.0,.3}
\newcommand{\red}{}
\newcommand{\blue}{}
\newcommand{\ps}{}
\newcommand{\od}{}

\newcommand{\cm}{}
\newcommand{\scrap}[1]{{\color{grey}{\sout{#1}}}}

\begin{document}
\title {Out-of-equilibrium quantum dot coupled to a microwave cavity}
\author{Olesia Dmytruk}
\affiliation{Laboratoire de Physique des Solides, CNRS UMR-8502, Universit\'{e} Paris Sud, 91405 Orsay cedex, France}
\author{Mircea Trif}
\affiliation{Laboratoire de Physique des Solides, CNRS UMR-8502, Universit\'{e} Paris Sud, 91405 Orsay cedex, France}
\author{Christophe Mora}
\affiliation{Laboratoire Pierre Aigrain,  Ecole Normale Superieure,
Université Paris 7 Diderot, CNRS; 24 rue Lhomond, 75005 Paris, France}
\author{Pascal Simon}
\affiliation{Laboratoire de Physique des Solides, CNRS UMR-8502, Universit\'{e} Paris Sud, 91405 Orsay cedex, France}
\date{\today}

\begin{abstract}
We consider a superconducting microwave cavity capacitively coupled to both a quantum conductor and its electronic reservoirs. We 
analyze  in details how the measurements of the cavity microwave field, which are related to the electronic charge susceptibility, can be  used to extract information on the transport properties of the quantum conductor.
We show that  the asymmetry of the capacitive couplings between the electronic reservoirs and the cavity plays a crucial role in relating 
optical measurements to transport properties. For asymmetric capacitive couplings, photonic measurements can be used to probe the finite  low frequency admittance of the quantum conductor, the real part of which being related to the differential conductance.
In particular, when  the quantum dot is far from resonance, the charge susceptibility is directly proportional to the admittance for a large range of frequencies and voltages. However, when the quantum conductor is near a resonance, such a relation generally holds only at low frequency and for equal tunnel coupling or low voltage. Beyond this low-energy near equilibrium regime, the charge susceptibility and thus the optical transmission offers new insights on the  quantum conductors since the optical observables are not directly connected to  transport quantities.
 For symmetric lead capacitive couplings, we show that the optical measurements can be used to reveal the Korringa-Shiba relation, connecting the reactive to the dissipative part of the susceptibility, at low frequency and low bias.
\end{abstract}

\pacs{74.78.Na, 74.25.N-, 42.50.Pq, 78.20.Bh}

\maketitle

\section{Introduction}

The charges transmitted through a mesoscopic conductor locally perturb the electromagnetic field.\cite{Hofheinz2011}Therefore, embedding the conductor in a high-finesse cavity offers a way to monitor charge transfer mechanisms with unprecedented accuracy~\cite{Schoelkopf1998} by coupling electronic transport with well-defined and controllable electromagnetic modes. \red{In the last decade, significant progress was made in engineering sensitive superconducting resonators~\cite{Wallraff2004,Paik2011} in the context of circuit Quantum Electrodynamics (QED).\cite{Blais2004}} Photons in these resonators have been coherently coupled to macroscopic two-level qubit systems mimicking quantum optics with strong atom-photon coupling~\cite{Fink2008,Niemczyk2010}. The strong reduction in the number of relevant degrees of freedom may allow to experimentally test the foundations of quantum physics~\cite{murch2013,xiang2013} and to implement quantum information~\cite{bergeal2010,menzel2012,devoret2013,yin2013,flurin2015} or quantum computation \cite{DiCarlo2009,mariantoni2011,fedorov2011} protocols.

\blue{
Superconducting strip-line resonators have been first coupled to isolated mesoscopic systems such as sets of 
isolated rings
\cite{reulet1995dynamic,deblock2002ac,deblock2002diamagnetic} as a probe to investigate the susceptibility of these systems without any connection to invasive probe.}
Superconducting resonators can also be coupled to Josephson junctions in order to build a qubit system but also  to mesoscopic open conductors such as quantum dots.\cite{childress2004,cottet-mora,BergenfeldtSET2012,Souquet2014,schiro-lehur,cottet}
Quantum dots are nanoscopic systems that emulate atoms by showing discrete energy levels, \red{atomic-like electronic filling shell structure} and being measurable by optical means. Contrary to atoms, they can be easily probed in electronic transport experiments similarly to large metallic conductors.  Combining resonant circuit QED with a quantum dot gives
a very powerful technique that allows to study the latter by photonic transport measurements~\cite{altimiras2014}
in addition to electronic transport. Recent experiments probe and manipulate single or double quantum dot (DQD) with resonators,\cite{delbecq2011,Petersson2012,Schroer2012,frey2012,toida2013,basset2013,delbecq2013,petta2013,viennot2014,viennot2015,basset2015} as well as demonstrate a DQD micromaser driven by 
single-electron tunneling.\cite{Lasing2011,petta2014,petta2015}

\begin{figure}[t] 
\centering
\includegraphics[width=0.7\linewidth]{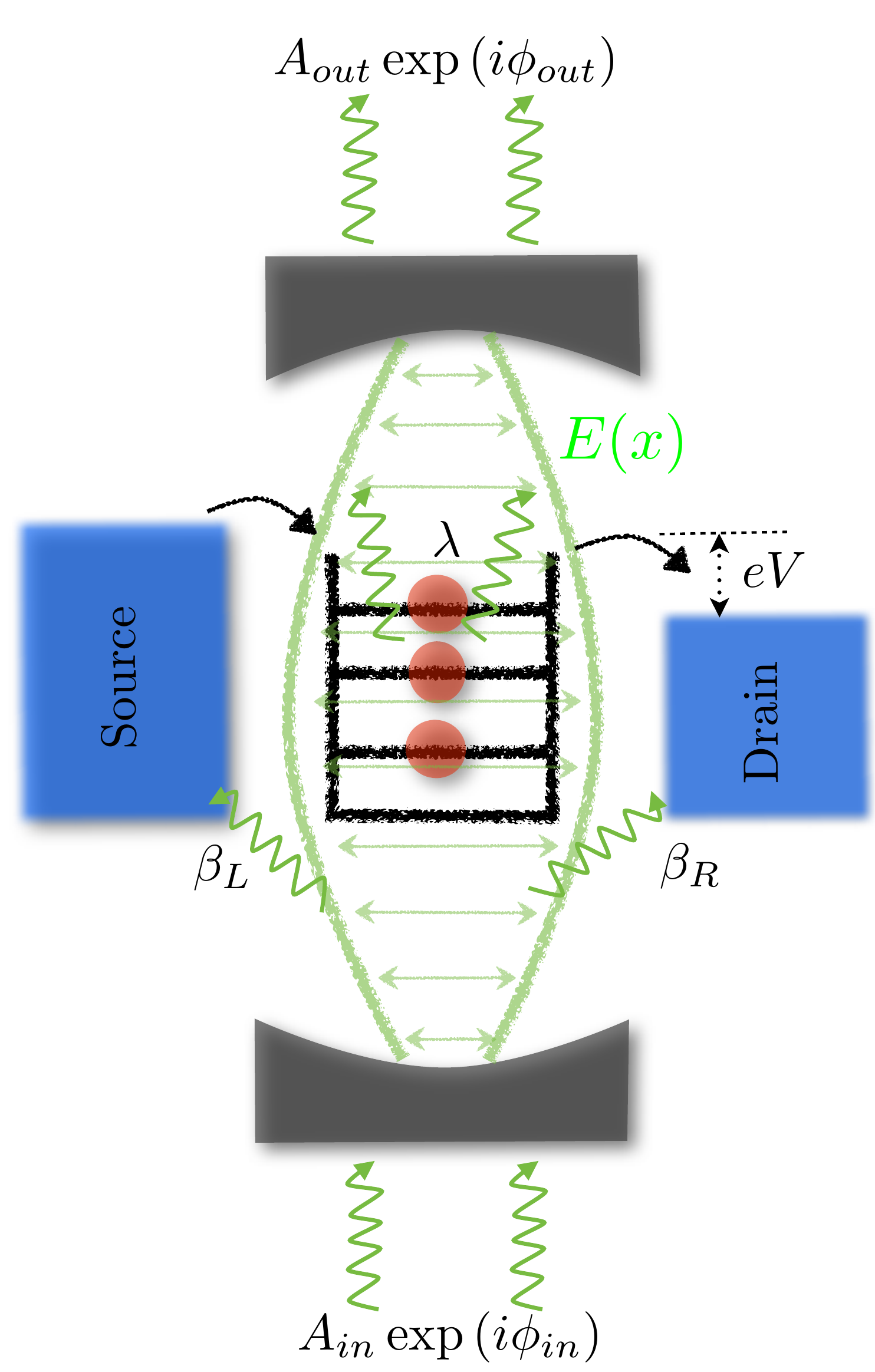}
\caption{Sketch of the system. The quantum dot with discrete energy levels (in black) is coupled to the source and drain electrodes (blue), which have a voltage bias $eV$. Differential conductance of the system may be found by transport measurements. The entire setup is coupled to the microwave cavity (depicted as light between two mirrors; mirrors in gray). The photons are coupled to the left lead with coupling strength $\beta_L$, to the right lead with $\beta_R$ and to the dot with $\lambda$. \red{The electromagnetic field inside the cavity is probed by sending input field with amplitude $A_{in}$ and phase $\phi_{in}$ and measuring the output field with $A_{out}$ and $\phi_{out}$. The difference between the outgoing and ingoing fields gives direct access to the charge susceptibility of the electronic system (see text).}} 
\label{Fig1} 
\end{figure}

 The photonic transport is quantified by the complex transmission coefficient  $\tau = A\exp(i\phi)$ that relates the output photonic field to the input field as shown in Fig.~\ref{Fig1}. Only the photons with the frequency $\omega$ near the cavity resonance frequency $\omega_c$ are \red{effectively} transferred across the microwave cavity.
\blue{
For isolated mesoscopic systems, it was shown that the measure of the  optical reflexion \od{coefficient} gives access to a cavity resonance frequency shift and  resonance frequency broadening which were related to the complex ac admittance.\cite{reulet1995dynamic}} For open mesoscopic conductors, such as quantum dots, the transmission coefficient at energy near the cavity frequency 
is sensitive to the charge susceptibility of the quantum dot circuit as was first proposed  in [\onlinecite{cottet-mora}].
\blue{Using the input-output formalism,\cite{walls2007} the transmission coeficient  was shown to read (see appendix~\ref{appendix_io})}:\cite{cottet-mora,simon}
\begin{align}
\tau(\omega) \approx \frac{\displaystyle \kappa}{\displaystyle -i(\omega - \omega_c) +  \kappa \od{+ i\,\Pi(\omega_c)}}\,.
\label{transmissioncoef}
\end{align}
where $\Pi(\omega)$ is proportional to the  electronic charge susceptibility. In this work, we assume that the perturbation of the cavity induced by the electrons is weak: $\od{|\Pi(\omega_c)}|\ll \kappa$, where $\kappa$ is the escape rate of the cavity that also controls the \red{frequency width} of photon resonant scattering in Eq.~\eqref{transmissioncoef}. Therefore,
we focus on the weak electron-photon coupling regime \red{where the electronic conductor only weakly perturbs the cavity}.

The reduction of the output signal compared to the input signal gives information about the imaginary part of the electronic susceptibility while the phase shift gives information about its real part, \blue{similar to what was found for  isolated mesoscopic systems.\cite{reulet1995dynamic}}
In what follows, we define the real (imaginary)  part of the susceptibility by $\Pi'(\omega)={\mathcal Re}[\Pi(\omega)]$ ($\Pi''(\omega)={\mathcal Im}[\Pi(\omega)]$).
More specifically, for $\omega\approx\omega_c$ and $|\Pi(\omega)|\ll \kappa$, we can approximate the phase shift as
\begin{align}
\delta\phi=\phi_{out}-\phi_{in}\approx \frac{\Pi'(\omega)}{\kappa},
\label{relation_phi}
\end{align}
which corresponds to the reactive part. Similarly, the cavity peak broadening translates into 
\begin{align}
\frac{\delta A}{A_{in}}=\frac{A_{in}-A_{out}}{A_{in}}\approx\frac{\Pi''(\omega)}{\kappa}\,,
\label{relation_A}
\end{align}
which corresponds to the dissipative part.

Most theoretical studies, excepting~ Ref.~[\onlinecite{cottet}] \cm{and} \od{[\onlinecite{benatov2012nanomechanical}]}, describe the coupling between the cavity and the quantum dot as the capacitive coupling between the cavity and the local charge on the dot. In this paper, we present a more realistic approach by including the capacitive coupling between the cavity and the metallic leads. We find that, in addition to being sensitive to charge fluctuations on the quantum dot, the cavity also probes charges that are directly transferred between the two leads. An unbalanced coupling between the cavity and the two leads is required to observe this charge transfer.
We evaluate the function $\Pi(\omega)$ for two systems: (i) a tunnel junction placed in a microwave cavity and (ii) a quantum dot (QD) tunnel coupled to metallic leads and placed in a microwave cavity, as showed schematically in Fig.~\ref{Fig1}. Although our model of the quantum dot does not incorporate electron interactions, we compare our calculations to the experimental results obtained by Delbecq {\it et al.} \cite{delbecq2011} especially in the near-equilibrium limit.

\ps{Our paper is organized as follows. Sec.~II describes our model Hamiltonian of the system presented in Fig.~\ref{Fig1}. In Sec.~III, we derive some general relations for the electronic charge susceptibility $\Pi$ and express it in terms of integrals over elements of the S-matrix for a non-interacting quantum conductor. First, we apply our formalism to the tunnel junction in Sec.~IV.  Then, we analyze the charge susceptibility of a non-interacting quantum dot coupled to a microwave stripline \red{cavity} in Sec.~V. \cm{The reader interested by the main findings of our work shall find them summarized in Sec.~VI.} 
Finally, we provide a short conclusion and a brief outlook in Sec.~VII.}

\section{Model Hamiltonian} 
Our system, depicted in Fig.~\ref{Fig1}, consists of a  quantum dot that is tunnel coupled to metallic leads. Both the quantum dot and the leads are capacitively coupled to a (superconducting) microwave cavity.  We assume that the wavelength of the electromagnetic field ($\sim$mm) is much larger than the mesoscopic system ($\sim$$\mu$m), and thus the coupling is constant over its length. The cavity supports many (in principle, infinitely many) modes, all of which should couple to the system. However, only \red{cavity} modes resonant with \red{some} electronic modes are relevant, the others giving negligible effect. From here on, we assume that only one mode, e.g. the lowest, couples to the system.  Nevertheless, our analysis can be straightforwardly extended to a multimode cavity. In this paper, we analyze the optical transmission and its relation to various observables of the mesoscopic system. We consider both the equilibrium and out-of-equilibrium situations where the voltage bias is large. 

The Hamiltonian of the combined system reads: 
\begin{align}\label{model-hamil}
H_{sys} = H_{el} + H_{el-c} + H_{ph}\,,
\end{align}

\begin{align}
H_{el} = \epsilon_d d^\dag d+ \sum_{k\alpha}\left(t_{\alpha}c_{k\alpha}^\dag d+h.c.\right)+\sum_{k\alpha} \epsilon_{k\alpha}c_{k\alpha}^\dag c_{k\alpha}\,,
\end{align}
\begin{align}
H_{el-c} = \lambda (a + a^\dag)d^\dag d + (a + a^\dag)\sum_{k\alpha}\beta_{\alpha}c_{k\alpha}^\dag c_{k\alpha}\,,
\end{align}
and $H_{ph} = \omega_c a^\dag a$. In these equations, $c_{k\alpha}$ ($c_{k\alpha}^\dagger$) is the annihilation (creation) operator of an electron in the left (right) lead $\alpha= L (R)$ with momentum $k$; $d$ ($d^\dagger$) is the annihilation (creation) operator of an electron in the quantum dot; $a$ ($a^\dagger$) is the annihilation (creation) operator of the photon field. 
We denote the energy level of the quantum dot as $\epsilon_d$, the tunneling amplitude between the quantum dot and the lead $\alpha$ as $t_\alpha$, and the cavity mode frequency as $\omega_c$. Finally, $\lambda$ and $\beta_\alpha$ are the electron-photon coupling constants to the quantum dot, and the lead $\alpha=L, R$, respectively. Note that we include the capacitive coupling between the quantum dot and the cavity as well as between the cavity and the electronic reservoirs.  
 \od{We have neglected  direct photon-induced tunneling
terms in the Hamiltonian which are subleading compared to capacitive couplings terms~\cm{\cite{cottet}. A general expression for the coupling constants  $\lambda$ and $\beta_\alpha$ is given in Ref. [\onlinecite{cottet}], involving an integral over the spatial variation of the vector potential in the corresponding part of the conductor, lead or quantum dot. The result depends on the geometry of the device and there is no general argument to favor  $\lambda$ over  $\beta_\alpha$ (or the opposite) in the general case.}


\section{Electronic charge susceptibility}\label{sec:susc}

\subsection{General relations obeyed by transport quantities}
According to Eq.~\eqref{transmissioncoef}, the weak coupling between the cavity and the electronic system results in  the modification of the transmission coefficient $\tau$ in the linear response regime\cite{simon} (see appendix~\ref{appendix_io} for a derivation of $\tau$).
In the time domain, the correlation function $\Pi(\omega)$ in Eq.~\eqref{transmissioncoef} reads
\begin{align}
\Pi(t - t') = &-i\theta(t - t')\langle\Big[\left(\lambda n_d + \beta_L n_L + \beta_R n_R\right)\left(t\right),\nonumber\\
\!\!&\left(\lambda n_d + \beta_L n_L + \beta_R n_R\right)\left(t'\right)\Big]\rangle,
\label{suscepttime}
\end{align}
which corresponds to the total charge susceptibility of the electronic system capacitively coupled to the cavity, where $n_d = d^\dag d$ is the number operator for electrons in the dot and $n_{\alpha} = \sum_{k} c^\dag_{k\alpha} c_{k\alpha}$ is the number operator of electrons in the lead $\alpha$. Note that we set $\hbar=1$ throughout the whole paper.

Using the fact that total charge is conserved,
$n_{d} + n_{L} + n_{R} = N,$
with $N$ proportional to the total charge operator  commuting with $H_{sys}$,
the susceptibility in Eq.(\ref{suscepttime}) can be rewritten as
\begin{align}
&\Pi\left(t - t'\right) = \left(\beta_L - \lambda\right)^2\Pi_{LL}\left(t - t'\right) + \left(\beta_R - \lambda\right)^2\Pi_{RR}\left(t - t'\right)\nonumber\\
\!\!&+\left(\beta_L - \lambda\right)\left(\beta_R - \lambda\right)\left[\Pi_{LR}\left(t - t'\right) + \Pi_{RL}\left(t - t'\right)\right],
\label{totalcorrfunc}
\end{align}
where we introduced 
\begin{align}
\Pi_{\alpha\beta}\left(t - t'\right) = -i\theta(t - t')\langle\left[n_\alpha\left(t\right), n_\beta\left(t'\right)\right]\rangle.
\label{corrfunc}
\end{align}
Using the equation of motion techniques, we can write
\begin{align}
&i\partial_t\Pi_{\alpha\beta}\left(t - t'\right) = \delta(t-t') \langle\left[n_\alpha\left(t\right), n_\beta\left(t'\right)\right]\rangle\nonumber\\
\!\!&+ \theta(t-t') \langle\left[I_\alpha\left(t\right), n_\beta\left(t'\right)\right]\rangle/e,
\label{eqofmotion1}
\end{align}
where
\begin{align}
I_{\alpha} = e\frac{\displaystyle dn_\alpha\left(t\right)}{\displaystyle dt}
\label{currentoperator}
\end{align}
is the current operator.
Let us introduce the correlation function 
\begin{align}
Y_{\alpha\beta}(t, t') = -i\theta(t-t')\langle[I_\alpha(t), n_\beta(t')]\rangle,
\label{diffconductance}
\end{align}
which may be interpreted as an admittance or as a generalized non-equilibrium differential conductance (see {\it e.g.}~[\onlinecite{safi}] with an opposite sign in the definition). Note that $Y$ depends on the voltage bias $V$ although it is not immediately visible from Eq. (\ref{diffconductance}).
Using Eq. (\ref{eqofmotion1}) and Eq. (\ref{diffconductance}) we  find 
the following relations between the susceptibility and the generalized differential conductance in Fourier space\footnote{We define the Fourier transform as
$
f(\omega) = \int_{-\infty}^{\infty}dt\exp(i\omega t) f(t),$}:
\begin{align}
\Pi_{\alpha\alpha}(\omega) = \frac{\displaystyle i}{\displaystyle e\omega}Y_{\alpha\alpha}(\omega),
\label{relsusceptcond1}
\end{align}
\begin{align}
\Pi_{LR}(\omega)+\Pi_{RL}(\omega) = \frac{\displaystyle i}{\displaystyle e\omega}\left[Y_{LR}(\omega)+Y_{RL}(\omega)\right].
\label{relsusceptcond2}
\end{align}
From Eqs (\ref{relsusceptcond1}) and  (\ref{relsusceptcond2}),  we can relate $\Pi(\omega)$ defined in Eq. (\ref{suscepttime}) to a linear combination of the admittances.
Such linear combination may be simplified in some limiting case as follows. For example, let us assume that there is no charge fluctuation in the quantum dot
so that  $n_d(t)\approx const$ in the cavity frequency range of interest. In this limit,
\begin{align}
\Pi(\omega) \approx \frac{\displaystyle i}{\displaystyle e\omega}(\beta_L-\beta_R)^2 Y(\omega),
\label{relsusceptcond_gen}
\end{align} 
where $Y=Y_{LL}=Y_{RR}=-Y_{LR}=-Y_{RL}$. $\Pi(\omega)$ is non-zero only for asymmetric couplings to the leads, $\beta_L\ne\beta_R$. 
The prediction of Eq.~(\ref{relsusceptcond_gen}) is quite intuitive: the cavity electric field shakes the electrons in the reservoirs asymmetrically and \red{therefore} probes the admittance of the whole system. 
Eq.~(\ref{relsusceptcond_gen}) also implies that the imaginary part of $\Pi(\omega)$ measured via the cavity resonance broadening is proportional  to the real part of the admittance, which, at low frequency, reduces to the 
 differential conductance. Thus the charge freezing limit $n_d(t)\approx const$ is a particular case where the optical transmission coefficient directly measures the non-equilibrium admittance of the system. Charge freezing can be obtained by replacing the quantum dot by an insulating barrier, by tuning the dot level far \red{off-resonance}, or by considering an interacting quantum dot in the Kondo regime as in Ref.~[\onlinecite{delbecq2011}]. We discuss this situation at the end of  Sec.~\ref{sec:dot}.

Introducing the non-symmetrized noise at finite frequency at
\begin{align}
C_{\alpha\beta}(\omega) = \int_{-\infty}^{\infty}dt\exp(i\omega t)\langle I_\alpha(t)I_\beta(0)\rangle,
\label{noise1}
\end{align}
the admittance $Y_{\alpha\alpha}(\omega)$ can be related to the current-current correlator \cite{safi}
through 
\begin{align}
C_{\alpha\alpha}(\omega) - C_{\alpha\alpha}(-\omega) = -2\omega e Y'_{\alpha\alpha}(\omega).
\label{currentcurrentcor3a}
\end{align}
The first term on the left hand side of Eq. (\ref{currentcurrentcor3a}) corresponds to the emission noise and the second term corresponds to the absorption noise.
%
%
Hence Eq.~(\ref{currentcurrentcor3a}) directly relates the increase in the cavity resonance broadening to the balance between the emission and absorption noise resulting from photon exchange between the quantum conductor and the 
cavity.\cite{mendes2015}

\subsection{Calculation of the electronic susceptibility}
In order to calculate the components of the total charge susceptibility introduced in Eq.~(\ref{corrfunc}), we use the scattering formalism that is well suited for such non-interacting problems.
In this formalism, the current operator $I_\alpha(t)$ \cite{blanter,rothstein} can be written as
\begin{align}
I_\alpha(t) = &\frac{\displaystyle e}{\displaystyle 2\pi}\int_{-\infty}^{\infty}dE\int_{-\infty}^{\infty}dE'\exp\left(i(E - E')t\right)\nonumber\\
&\times\sum_{\gamma \gamma '}A_{\gamma \gamma '}(\alpha, E, E')a^\dag_\gamma(E)a_{\gamma '}(E'),
\label{definitioncurrent}
\end{align}
where
\begin{align}
A_{\gamma \gamma '}(\alpha, E, E ') = \delta_{\alpha \gamma '}\delta_{\alpha \gamma} - S^*_{\alpha \gamma}(E)S_{\alpha \gamma '}(E').
\label{A}
\end{align}
In Eq.~(\ref{A}),  $S_{\alpha \gamma}(E)$ are the elements of the scattering matrix characterizing the system (tunnel junction, quantum dot, etc.). The current operator $I_\alpha$ is expressed in terms of annihilation (creation) $a_\alpha(E)$ ($a^\dag_\alpha(E)$) operators of the electrons in the reservoir connected to terminal $\alpha$. These operators are normalized so that
\begin{align}
\langle a^\dag_\alpha(E)a_{\alpha '}(E ')\rangle = \delta_{\alpha\alpha '}\delta(E - E ')f_\alpha(E),
\label{normalization}
\end{align}
where 
$f_\alpha(E) = [\exp(E - \mu_\alpha)/k_BT + 1]^{-1}$ denotes the Fermi function in the lead $\alpha$ and $\mu_\alpha$ is its chemical potential.

The Fourier transform of $I_\alpha(t)$ reads:
\begin{align}
I_\alpha(\omega) = e\int_{-\infty}^{\infty}dE\sum_{\gamma \gamma '}A_{\gamma \gamma '}(\alpha, E, E + \omega)a^\dag_\gamma(E)a_{\gamma '}(E + \omega).
\label{currentoperator1}
\end{align}

Using 
\begin{align}
n_\alpha(t) = \frac{\displaystyle i}{\displaystyle 2\pi e}\int_{-\infty}^{\infty}d\omega\exp(-i\omega t) \frac{\displaystyle I_\alpha(\omega)}{\displaystyle \omega}
\end{align}
and Eq. (\ref{currentoperator1}), we can express the number operator as
\begin{align}
n_\alpha(t)
& =\frac{\displaystyle i}{\displaystyle 2\pi}\int_{-\infty}^{\infty}d\omega\exp(-i\omega t) \frac{\displaystyle 1}{\displaystyle \omega}
\label{densityoperator}\\
&\times\int_{-\infty}^{\infty}dE\sum_{\gamma \gamma '}A_{\gamma \gamma '}(\alpha, E, E + \omega)a^\dag_\gamma(E)a_{\gamma '}(E + \omega).\nonumber
\end{align}

Performing manipulations detailed in the appendix \ref{appendix_smatrix}, we can rewrite $\Pi_{\alpha\beta}$ as
\begin{align}
&\Pi_{\alpha\beta}(\omega) = \left(\frac{\displaystyle 1}{\displaystyle 2\pi}\right)^2\int_{-\infty}^{\infty}\int_{-\infty}^{\infty}
d\omega_2dE_2\Bigg\{
\frac{\displaystyle 1}{\displaystyle \omega_2^2}\frac{\displaystyle 1}{\displaystyle \omega + \omega_2 + i\eta}\nonumber\\
&\times \sum_{\gamma_1,\gamma'_1}F^{\alpha\beta}_{\gamma_1\gamma'_1}(E_2, \omega_2)\left[f_{\gamma_1}(E_2 + \omega_2) - f_{\gamma'_1}(E_2)\right]\Bigg\},
\label{corrfuncresult2}
\end{align}

where we introduced a new function
\begin{align}
F^{\alpha\beta}_{\gamma\gamma'} = A_{\gamma\gamma'}(\alpha, E + \omega, E)A_{\gamma'\gamma}(\beta, E, E + \omega).
\label{fgeneral}
\end{align}

At zero temperature ($T = 0$), this expression can be simplified further.
Introducing the integrals
\begin{align}
K_{\alpha\beta}(\omega_2) = &\int_{eV/2}^{eV/2 - \omega_2}dE_2
F^{\alpha\beta}_{LL}(E_2, \omega_2)\nonumber\\
&+ \int_{-eV/2}^{-eV/2 - \omega_2}dE_2
F^{\alpha\beta}_{RR}(E_2, \omega_2)\nonumber\\
&+ \int_{-eV/2}^{eV/2 - \omega_2}dE_2
F^{\alpha\beta}_{LR}(E_2, \omega_2)\nonumber\\
&+ \int_{eV/2}^{-eV/2 - \omega_2}dE_2
F^{\alpha\beta}_{RL}(E_2, \omega_2),
\label{kalphabeta}
\end{align}
the real and imaginary part of the susceptibility can be written as
\begin{align}
\Pi'_{\alpha\beta}(\omega) = \left(\frac{\displaystyle 1}{\displaystyle 2\pi}\right)^2\mathcal{P}\int_{-\infty}^{\infty}d\omega_2\frac{\displaystyle 1}{\displaystyle \omega_2^2}\frac{1}{\omega + \omega_2}K_{\alpha\beta}(\omega_2),
\label{palphabeta}
\end{align}

\begin{align}
\Pi''_{\alpha\beta}(\omega) = -\frac{\displaystyle \pi}{\displaystyle \left(2\pi\omega\right)^2}K_{\alpha\beta}(-\omega).
\label{dalphabeta}
\end{align}
In order to compute the susceptibility $\Pi_{\alpha\beta}$ for a non-interacting scatterer in question, we first need to compute the functions $F^{\alpha\beta}_{\gamma\gamma'}$ in Eq.~(\ref{fgeneral}), and then the functions $K_{\alpha\beta}$ in Eq.~(\ref{kalphabeta}). In the next section, we present these computations for a simpler system~-- the tunnel junction,~-- and we analyze a more complex quantum dot system in Sec. V.

\section{Tunnel junction}

\subsection{Cavity pull and dissipation}

The Hamiltonian that combines the resonating cavity with the tunnel junction has the form of Eq.~\eqref{model-hamil} with 
\begin{equation}
H_{el} = \sum_{k\alpha} \epsilon_{k\alpha}c_{k\alpha}^\dag c_{k\alpha}  + \sum_{k q}\left(t c_{k R}^\dag c_{qL}+h.c.\right) 
\end{equation}
and
\begin{equation}\label{elc}
H_{el-c} =  (a + a^\dag)\sum_{k\alpha}\beta_{\alpha}c_{k\alpha}^\dag c_{k\alpha}\,,
\end{equation}
characterized by the capacitive coupling $\beta_L$ ($\beta_R$) of the left (right) lead to the cavity. The tunneling amplitude $t$ being independent of energy is equivalent to instantaneous electron scattering. As a result, no charge can accumulate in the junction, the total charge in the leads $N=n_R+n_L$ is constant, and the current is conserved $I_R = - I_L$. The different charge 
susceptibilities are related to each other 
\begin{equation}\label{current-conservation}
\Pi_{LL}(\omega) = \Pi_{RR}(\omega) = -\Pi_{LR}(\omega) = -\Pi_{RL}(\omega)
\end{equation}
so that the total electronic susceptibility in Eq.~\eqref{transmissioncoef} reads
\begin{equation}
\Pi(\omega) = \sum_{\alpha,\gamma} \beta_\alpha \beta_\gamma \Pi_{\alpha \gamma} (\omega) = (\beta_L - \beta_R)^2 \Pi_{LL}(\omega),
\end{equation}
as in Eq.~\eqref{relsusceptcond_gen}. The dependence on the coupling difference $\beta_L - \beta_R$ may be also viewed as a consequence of gauge invariance as discussed in Sec.~\ref{gauge}.

Using the scattering formalism of Sec.~\ref{sec:susc}, we can compute the electronic susceptibility for the tunnel junction. The scattering matrix is energy independent and takes the form
\begin{align}
S = \begin{pmatrix}
ir & t \\
 t & ir
\end{pmatrix},
\label{scattmatrixtj}
\end{align}
where unitarity implies $r^2 + t^2 = 1$ (we choose real parameters for simplicity). The functions introduced in Eq.~\eqref{fgeneral} are then expressed as $F_{LL}^{\alpha \beta} = F_{RR}^{\alpha \beta} = \varepsilon_{\alpha \beta} T^2$ and $F_{LR}^{\alpha \beta} = F_{RL}^{\alpha \beta} = \varepsilon_{\alpha \beta} T (1-T),$ where $T = t^2$ is the tunnel junction transmission, $\varepsilon_{\alpha \beta} = 1$ if $\alpha=\beta$ and $-1$ otherwise. Inserting these expressions in Eq.~\eqref{kalphabeta}, we obtain
\begin{equation}
K_{\alpha\beta}(\omega_2) = -2 T \omega_2 \, \varepsilon_{\alpha \beta}.
\end{equation}
Even at this level one may observe that the photonic transmission is not affected by the bias voltage $V$. This is due to the linearity in the $I$-$V$ characteristic of the tunnel junction.

Integrating Eq.~\eqref{palphabeta}
\begin{equation}
P \int_{-\infty}^{\infty} \frac{d \omega_2}{\omega_2 (\omega + \omega_2)} = 0,
\end{equation}
we find that the real part of the susceptibility becomes zero, $\Pi'_{\alpha \beta} = 0$, hence the absence of the cavity pull. This result is in agreement with Ref.~\onlinecite{mendes2015} where a different gauge was used, corresponding to the 
Eq.~\eqref{second-gauge} derived in the next subsection. Therefore, any shift in the cavity resonant frequency is due to either a weak energy dependence in the electronic transmission or to higher order effects (backaction) in the electron-photon coupling. The imaginary part is obtained from Eq.~\eqref{dalphabeta} and reads $\Pi''_{\alpha \beta} (\omega) = - \left(T/2 \pi \omega\right) \varepsilon_{\alpha \beta}$ \red{recovering} the current conservation in Eq.~\eqref{current-conservation}. Finally, the total susceptibility of the tunnel junction
\begin{equation}\label{suscep-final}
\Pi(\omega) =  - i (\beta_L - \beta_R)^2 \frac{T}{2 \pi \omega}
\end{equation}
corresponds to a purely dissipative effect of the electrons on the photonic transmission. This expression can be also obtained by computing the admittance $Y(\omega)$ and using Eq.~\eqref{relsusceptcond_gen}. For the tunnel junction, the admittance  $Y(\omega)$ does not depend on neither the frequency $\omega$ nor the voltage $V$. It is equal to the linear conductance $G = \frac{\partial I}{\partial V}$. 

\subsection{Gauge invariance}\label{gauge}
Different choices of gauge are possible to describe the same physical system. They are related to each other via unitary transformations. In order to relate to the previous work, we discuss below two other gauges.

The first unitary transformation is given by the displacement operator $U_1 = e^{(\beta_R/\omega_c)  n (a^\dag-a)}$, shifting the photon field $a \to a - (\beta_R/\omega_c) n$. The operator $U_1$ does not affect the dynamics of electrons since the total number of electrons $n= n_R+n_L$ commutes with the Hamiltonian, hence $U_1 H_{el} U_1^\dag =  H_{el}$. The operator $U_1$ substracts $\beta_R$ from the capacitive couplings thereby canceling the coupling to the right lead. More specifically, the transformed Hamiltonian is given by
\begin{align}
\nonumber U_1 (H_{ph} + & H_{el-c} ) U_1^\dag = \omega_c a^\dag a + (a+ a^\dag ) (\beta_L - \beta_R) n_L \\[2mm]
&- \frac{\beta_R \beta_L}{\omega_c} n^2 - \frac{\beta_R (\beta_L-\beta_R )}{\omega_c} n (n_L-n_R).
\label{gauge1}
\end{align}
The first two terms describe the single mode resonator and its coupling to the left lead. The second term particularly emphasizes the sensitivity of the photon transmission to the capacitive inhomogeneity $\beta_L-\beta_R$. Since $n$ commutes with the total Hamiltonian, the third term is a constant and the last term describes an additional small bias voltage. The result of Eq.~\eqref{suscep-final} may be obtained from Eq.~\eqref{gauge1} straightforwardly.

The second unitary transformation removes the linear coupling between the cavity and the electrons, i.e. Eq.~\eqref{elc}, by dressing the tunneling amplitude between the two leads. This transformation with the unitary operator $U_2 = e^{A (a^\dag - a)}$, with $\omega_c A = \beta_L n_L + \beta_R n_R$, transforms the Hamiltonian into
\begin{align}
\nonumber H' = & U_2 H U_2^\dag = \sum_{k q}\left(t e^{(\beta_R - \beta_L) (a^\dag - a)/\omega_c} c_{k R}^\dag c_{qL}+h.c.\right)  \\ 
& +\sum_{k\alpha} \epsilon_{k\alpha}c_{k\alpha}^\dag c_{k\alpha} + \omega_c a^\dag a - \frac{A^2}{\omega_c}. \label{second-gauge}
\end{align}
The last term in this expression  $A^2/\omega_c$ describes capacitive electron-electron interaction across the tunnel junction and is usually neglected.
The resulting Hamiltonian is conventionally invoked to describe the physics of dynamical Coulomb blockade~\cite{Hofheinz2011,Souquet2014,altimiras2014}.

We remark that the different gauges correspond to shifted positions of the resonator vacuum. A vacuum in one gauge is different from the vacuum in another gauge.


\section{Quantum dot}\label{sec:dot}
Let us determine the charge susceptibility probed by the microwave cavity of the QD tunnel coupled to metallic leads and capacitively coupled to the microwave cavity.
The scattering matrix for the single-level quantum dot reads 
\begin{align}
S(E) = -1 + ig(E)\begin{pmatrix}
\Gamma_L & \sqrt{\Gamma_L\Gamma_R} \\
 \sqrt{\Gamma_L\Gamma_R} & \Gamma_R
\end{pmatrix},
\label{scattmatrixqd}
\end{align}
where
\begin{align}
g(E) = \frac{\displaystyle 1}{\displaystyle E - \epsilon_d + i\Gamma/2},
\end{align}
and $\Gamma=\Gamma_L+\Gamma_R$ is the width of the level. Using this expression for the $S$-matrix we can compute
the functions $A_{\gamma\gamma'}(\alpha,E,E')$ defined in Eq.~\eqref{A}, then the sixteen functions  $F_{\gamma\gamma'}^{\alpha\beta}$ introduced in 
Eq.~\eqref{fgeneral}, and finally the functions $K_{\alpha\beta}(\omega)$ defined in Eq.~\eqref{kalphabeta}.
Being lengthy and not essential for further discussion, the intermediate expressions for the functions $F_{\gamma\gamma'}^{\alpha\beta}$  and  $K_{\alpha\beta}(\omega)$ are given in the Supplementary Material~\cite{SupMat}.
The functions $K_{\alpha\beta}$  directly provide $\Pi''_{\alpha\beta}$ through Eq.~\eqref{dalphabeta}. However, in order to obtain $\Pi'_{\alpha\beta}$, we need to perform a final tedious integration over energy in Eq.~\eqref{palphabeta},
the details of which are given in the Appendix~\ref{appendix_susc}.
The final expressions for the real and imaginary parts of the total susceptibility are given in Eq.~\eqref{eq:pi_r} and Eq.~\eqref{eq:pi_i}, respectively,
of the Appendix~\ref{appendix_susc}. However, such expressions are intricate and difficult to interpret directly. Therefore, we consider some limiting cases where simple expressions and conclusions can be obtained before addressing the general case.

\subsection{The Quantum RC-circuit limit }
For the two following limits (or sets of parameters) our system 
reduces to the so-called quantum RC circuit, which has been largely studied ~theoretically\cite{buttiker1993a,buttiker1993b,nigg2006,mora-lehur,hamamoto2010,filippone2011,lee2011,filippone2012,golub2012,filippone2013,mora2013,dutt2013,lee2014,cottet} 
as well as implemented and verified experimentally~\cite{gabelli2006,feve2007,gabelli2012}.

\subsubsection{Case $\Gamma_R=0$}
This case corresponds to a quantum dot connected to a single reservoir by a junction.
The quantum RC circuit is particularly interesting in the small frequency limit when some universal behavior is expected~\cite{nigg2006,mora-lehur}
and was observed~\cite{gabelli2006}.

When  $\omega$ is smaller than all other energy scales at play, we can further simplify the expression of the susceptibility $\Pi(\omega)$ to obtain
\begin{align}
\Pi'(\omega)=-\frac{\displaystyle 2\Gamma_L (\beta_L-\lambda)^2}{\displaystyle \pi\left(\Gamma_L^2+4\epsilon_d^2\right)}
+O(\omega^2)\,,
\end{align}
for the real part, and 
\begin{align}
\Pi''(\omega)=-\frac{\displaystyle 4\Gamma_L^2\omega(\beta_L-\lambda)^2}{\displaystyle \pi\left(\Gamma_L^2+4\epsilon_d^2\right)^2}+O(\omega^3)\,,
\end{align}
for the imaginary part. 
Note that these expressions correspond exactly to $\Delta\omega_0^a$ and $\Delta\Lambda_0$ in Ref.~\onlinecite{cottet} with $\Gamma=\Gamma_L/2$ [up to a minus sign which can be traced back to our different definition of $\Pi(t)$ in Eq.~\eqref{suscepttime}].

The real and imaginary contributions, $\Pi'$ and $\Pi''$, respectively,  are not independent
and \red{fulfill} at the lowest order in $\omega$  the  following relation
\begin{align}\label{korringa}
\Pi''(\omega)=-\pi \omega\left[\Pi'(0)\right]^2/ (\lambda-\beta_L)^2.
\end{align}
Up to the electron-photon couplings, this is nothing but the Korringa-Shiba relation connecting the dissipative part of the susceptibility to its reactive part. Note that we defined the correlation function $\Pi(t)$ in Eq.~\eqref{suscepttime} with a different sign compared to, e.g., Ref.~\onlinecite{mora-lehur}, and thus introduced the opposite sign in the Korringa-Shiba relation. Note that the use of a microwave cavity for verifying the Korringa Shiba relation has already been proposed in Ref. [\onlinecite{cottet}].

\subsubsection{Case $\beta_L=\beta_R$}
 For equal capacitive coupling to the leads ($\beta_L=\beta_R$), the cavity electric field effectively couples to a single larger lead as well as  to the quantum dot. We thus expect to observe the same physics as in the previous case.  Again, we assume $\omega$ to be \red{smaller} than the other energy scales in the system.

In this low-frequency limit, the real part of the total susceptibility $\Pi$ 
reads
\begin{align}
\Pi'(\omega)=&-\frac{\displaystyle 2\left(\lambda-\beta_L\right)^2}{\displaystyle \pi}\left[\frac{\displaystyle \Gamma_L}{\displaystyle \left(eV-2\epsilon_d\right)^2+\Gamma^2}\right.\nonumber\\
\!\!&\left.+\frac{\displaystyle \Gamma_R}{\displaystyle \left(eV+2\epsilon_d\right)^2+\Gamma^2}\right]+O(\omega^2)\,,
\label{realqdsmallomega}
\end{align}
while the imaginary part reads
\begin{align}
\Pi''(\omega)=&-\frac{\displaystyle 4\Gamma\omega\left(\lambda-\beta_L\right)^2}{\displaystyle \pi}\left[\frac{\displaystyle \Gamma_L}{\displaystyle \left(\left(eV-2\epsilon_d\right)^2+\Gamma^2\right)^2}\right.\nonumber\\
\!\!&+\left.\frac{\displaystyle \Gamma_R}{\displaystyle \left(\left(eV+2\epsilon_d\right)^2+\Gamma^2\right)^2}\right]
+O(\omega^3).
\label{imqdsmallomega}
\end{align}

For zero bias voltage $eV=0$, one recovers the Korringa-Shiba relation Eq.~\eqref{korringa}. Indeed, the two leads combine and the quantum dot effectively couples to a single lead.

In the latter case ($\beta_L=\beta_R$), given that $eV=0$, as well as in the former case ($\Gamma_R = 0, eV=0$), the cavity measures only local charge fluctuations on the dot~-- leading to the dissipation channel fulfilling the Korringa-Shiba relation,~--  but it is not sensitive to the charge transfer across the quantum dot.

\subsection{The tunneling limit, $\epsilon_d/\Gamma\rightarrow-\infty$}

The transmission coefficient for the quantum dot, $T_d$, in the tunneling limit ($\epsilon_d \rightarrow -\infty$) reads
\begin{align}
T_d=\frac{\displaystyle \Gamma_L\Gamma_R}{\displaystyle \epsilon_d^2}.
\end{align}
Using this equation, we can immediately evaluate the real and imaginary parts of $\Pi$:
\begin{align}
\Pi'(\omega)=&-\frac{\displaystyle T_d}{\displaystyle 2\pi}\left[\frac{(\beta_L-\lambda)^2}{\Gamma_R} +
\frac{(\beta_R-\lambda)^2}{\Gamma_L}+\right. \nonumber\\
&~~~~~~~~\left. (\beta_L-\lambda)(\beta_R-\lambda)\left(\frac{1}{\Gamma_L}+ \frac{1}{\Gamma_R}\right)T_d\right]\,,
\label{tunreal}
\end{align}
\begin{align}
\Pi''(\omega)=-\frac{\displaystyle T_d}{\displaystyle 2\pi\omega}(\beta_L-\beta_R)^2.
\label{tunim}
\end{align}

The imaginary part of the susceptibility of the QD, related to $\delta A/A$, Eq.~\eqref{tunim}, is equal to the susceptibility of the tunnel junction in Eq.~\eqref{suscep-final}. The real part of the susceptibility of the QD, related to the phase shift $\delta\phi$, is not zero  as opposed to the tunnel junction result. The non-zero phase shift for the QD is explained by the existence of a localized level inside the barrier which is coupled virtually to the continuum of levels in the leads by the cavity field. On the other hand, such level is absent in a pure tunnel junction resulting in the zero real part of the susceptibility.

\subsection{The low-frequency limit}

Let us explore in detail the low-frequency regime when the cavity frequency is typically smaller
than the electronic quantum dot resonance width, $\omega\ll\Gamma$. This regime is studied experimentally in 
[\onlinecite{delbecq2011}].

In this limit, one can take a series in $\omega/\Gamma$ of both the real and imaginary parts of the charge susceptibility: 
\begin{align}
\Pi'_{\alpha\beta}(\omega)&=\Pi'_{\alpha\beta}(0)+O(\omega^2),\\
\Pi''_{\alpha\beta}(\omega)&=\frac{1}{\pi\omega}\Lambda^{(-1)}_{\alpha\beta}+\omega\Lambda^{(1)}_{\alpha\beta}+O(\omega^3).
\end{align}

We obtain the following expressions for the lowest order terms.
\begin{align}
\Pi'_{LL,RR}(0)=-\frac{\displaystyle 2\Gamma_{L,R}\left[\left(eV\mp2\epsilon_d\right)^2\pm(\Gamma_L^2-\Gamma_R^2)\right]}{\displaystyle \pi\left[\left(eV\mp2\epsilon_d\right)^2+\Gamma^2\right]^2},
\end{align}
\begin{align}
&\Pi'_{LR}(0)+\Pi'_{RL}(0)=-\frac{\displaystyle 4\Gamma\Gamma_L\Gamma_R}{\displaystyle \pi}\times\nonumber\\
&\left[\frac{\displaystyle 1}{\displaystyle \left(\left(eV-2\epsilon_d\right)^2+\Gamma^2\right)^2}+\frac{\displaystyle 1}{\displaystyle \left(\left(eV+2\epsilon_d\right)^2+\Gamma^2\right)^2}\right].
\end{align}
and similarly

\begin{align}
\Lambda^{(-1)}_{LL,RR}=-\frac{\displaystyle 2\Gamma_L\Gamma_R}{\displaystyle \left[\left(eV\mp2\epsilon_d\right)^2+\Gamma^2\right]}\,,
\label{qdsmallomegall}
\end{align}

\begin{align}
&\Lambda^{(-1)}_{LR}+\Lambda^{(-1)}_{RL}=2\Gamma_L\Gamma_R\times\nonumber\\
\!\!&\times\left[\frac{\displaystyle 1}{\displaystyle \left(eV-2\epsilon_d\right)^2+\Gamma^2}+\frac{\displaystyle 1}{\displaystyle \left(eV+2\epsilon_d\right)^2+\Gamma^2}\right].
\label{qdsmallomegalrrl}
\end{align}

The expressions for  $\Lambda^{(1)}_{\alpha\beta}$  read:
\begin{align}
\Lambda^{(1)}_{LL,RR}=&-\displaystyle \frac{4\Gamma_{L,R}}{3\pi\left[\left(eV\mp2\epsilon_d\right)^2+\Gamma^2\right]^{3}}\times\nonumber\\
&
\hspace{-1.5cm}\left[(3\Gamma_{L,R}-2\Gamma_{R,L})\Gamma^2+3(\Gamma+\Gamma_{R,L})(eV\mp2\epsilon_d)^2\right],
\label{qdsmallomegall1}
\end{align}

and
\begin{align}
\Lambda^{(1)}_{LR}+\Lambda^{(1)}_{RL}=&
-\frac{\displaystyle 4\Gamma_L\Gamma_R}{\displaystyle3\pi}\left[\frac{\displaystyle 5\Gamma^2-3(eV-2\epsilon_d)^2}{\displaystyle \left[\left(eV-2\epsilon_d\right)^2+\Gamma^2\right]^3}\right.\nonumber\\
&+\left.\frac{\displaystyle 5\Gamma^2-3(eV+2\epsilon_d)^2}{\displaystyle \left[\left(eV+2\epsilon_d\right)^2+\Gamma^2\right]^3}\right].
\end{align}

Analyzing these equations, one may observe that both the real and imaginary parts of the susceptibility $\Pi_{\alpha\beta}(\omega)$ have peaks at $\epsilon_d=\pm eV/2$. 
Noticing that $\Lambda^{(-1)}_{LL}+\Lambda^{(-1)}_{RR}+\Lambda^{(-1)}_{LR}+\Lambda^{(-1)}_{RL}=0$, we can further simplify the expression for $\Pi''(\omega)$ as
\begin{align}
\Pi''(\omega)=&\frac{(\beta_L-\beta_R)}{\pi\omega}\left[(\beta_L-\lambda)\Lambda^{(-1)}_{LL}-(\beta_R-\lambda)\Lambda^{(-1)}_{RR}
\right]\nonumber\\
&+\omega\left[ (\beta_L-\lambda)^2\Lambda^{(1)}_{LL}+(\beta_R-\lambda)^2\Lambda^{(1)}_{RR}\right.\nonumber\\
&~~~+\left.(\beta_L-\lambda)(\beta_R-\lambda)
(\Lambda^{(1)}_{LR}+\Lambda^{(1)}_{RL}) \right]\,.
\end{align}
Let us discuss  both $\beta_L=\beta_R$ and $\beta_L\ne \beta_R$ cases.

\subsubsection{Case $\beta_L=\beta_R$}
For $\beta_L=\beta_R$, we find again the effective quantum RC circuit case discussed above. 
In particular, it implies that at $V=0$, 
\begin{equation}\label{deltaphi_lorentz}
\delta\phi(V=0)\approx- \frac{2\Gamma(\beta_L-\lambda)^2}{\pi\kappa\left(\Gamma^2+4\epsilon_d^2\right)},
\end{equation}
which corresponds to a Lorenztian shape as a function of the dot gate voltage $\propto\epsilon_d$. 

In order to analyze the general non-equilibrium  $V\ne 0$ case, we plot in Fig.~\ref{lowfsym} both $\tilde{\Pi}'=\Pi'/(\beta_L-\lambda)^2$  and $\tilde{\Pi}''=\Pi''/(\beta_L-\lambda)^2$ as functions of $\epsilon_d$ and $eV$.
\begin{figure}[t] 
\centering
\includegraphics[width=0.95\linewidth]{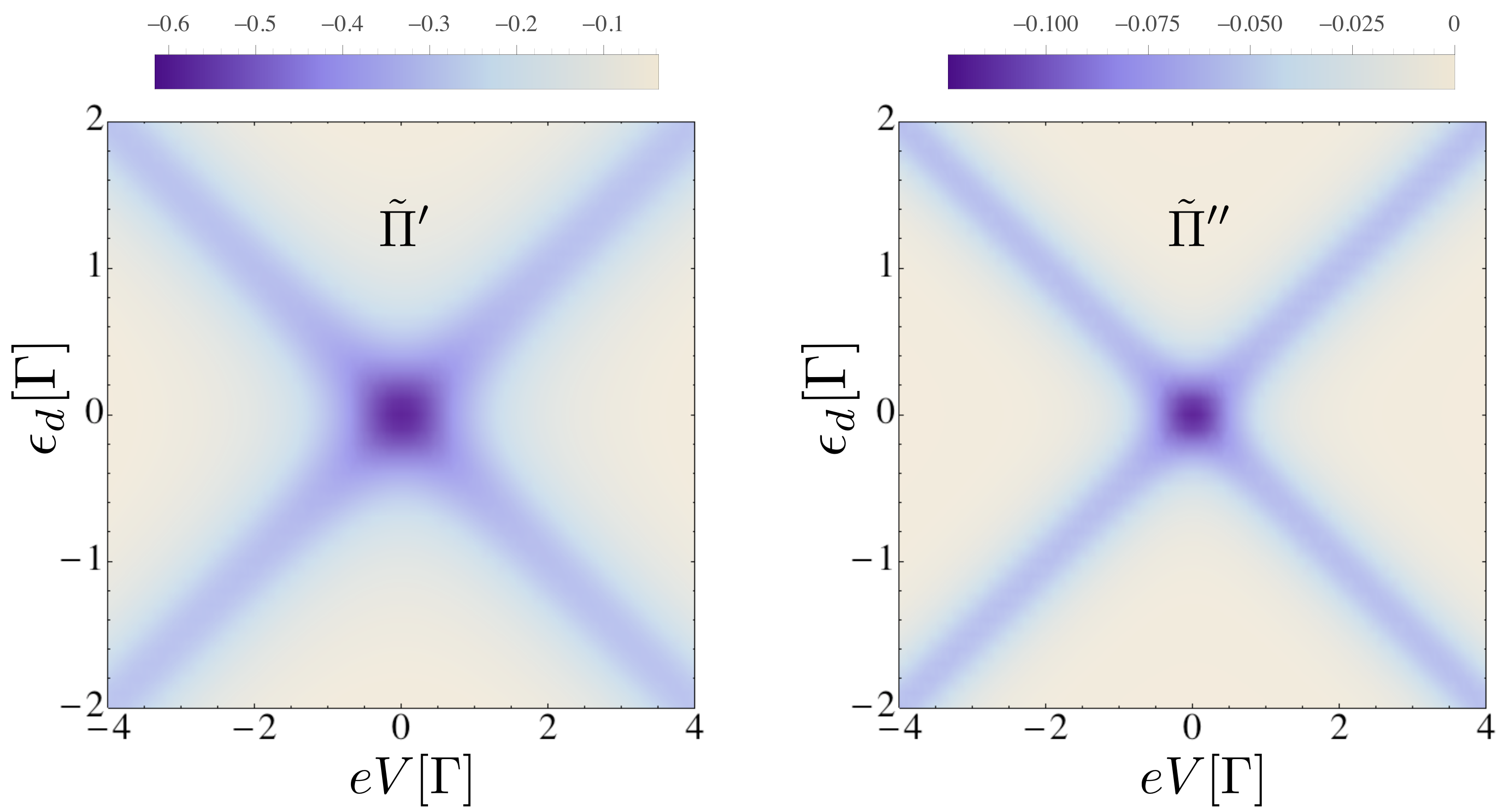}
\caption{$\tilde{\Pi}'(\omega)\equiv\Pi'(\omega)/\left(\beta_L-\lambda\right)^2$ (left) and $\tilde{\Pi}''(\omega)\equiv\Pi''/(\beta_L-\lambda)^2$ (right)
as a function of $eV$ and $\epsilon_d$ for $\beta_L=\beta_R$, $\Gamma_L=\Gamma_R=0.5$ and $\omega=0.1$. All energies are expressed in terms of $\Gamma=\Gamma_L+\Gamma_R$. } 
\label{lowfsym} 
\end{figure}
Both the real and imaginary parts of  $\Pi$ have peaks around $eV=0$. For $eV$ larger than $\Gamma$, the peaks split, and resonances are found for $eV/2=\pm \eps_d$. For $\beta_L=\beta_R$, the Korringa Shiba relations are satisfied at low energy for $V=0$. This implies that $\Pi'' \propto (\Pi')^2$ and therefore  the resonances in $\Pi''$ are more narrow at $eV=0$ than the resonances of $\Pi'$.

Let us compare these results to the non-equilibrium differential conductance $G(V)=dI/dV$ \red{across} the quantum dot that reads
\begin{equation}
G(V)= \frac{e^2}{2\pi} \left(\frac{2\Gamma_L\Gamma_R}{(eV-2\epsilon_d)^2+\Gamma^2}+\frac{2\Gamma_L\Gamma_R}{(eV+2\epsilon_d)^2+\Gamma^2}\right).
\end{equation}
The differential conductance has a Lorentzian shape double peaked at $eV/2=\pm \eps_d$. 
Noticing that, for $\Gamma_L=\Gamma_R$, the expression of $\Pi'(\omega,V)$ given in Eq.~\eqref{realqdsmallomega}
is directly proportional to $G(V)$. We thus expect that
\begin{equation} \label{eq:compare}
\delta\phi(V)=-\frac{4(\beta_L-\lambda)^2}{\kappa e^2 \Gamma} G(V),
\end{equation}
for $\Gamma_L=\Gamma_R=\Gamma/2$. 

Therefore, for both symmetric tunneling amplitudes and lead capacitive couplings, Eq. \eqref{eq:compare} directly relates the optical phase shift to the non-equilibrium conductance. This is one of our main results.

For small deviations from the equalities $\Gamma_L=\Gamma_R$ and $\beta_L=\beta_R$, we expect the relation Eq.~\eqref{eq:compare} to hold approximately. For strong asymmetric tunneling, even though the expressions for $\Pi'$ and $G(V)$ are no longer proportional within each other, these two quantities still have
a similar shape characterized by Lorentzian peaks around $eV/2=\pm \eps_d$. 

Such quantitative comparison cannot be made between $\Pi''$ (and therefore $\delta A/A$) and $G(V)$ for $\beta_L=\beta_R$ because the term proportional to $1/\omega$ in the series for $\Pi''(\omega)$ is exactly zero. However, small deviations of the stringent condition $\beta_L=\beta_R$ may allow comparing these quantities as we show next.

\subsubsection{Case $\beta_L\ne \beta_R$}

In the more general case $\beta_L\ne \beta_R$, the expression for $\Pi''$ simplifies for $eV=0$ as
\begin{align}
\!\!\!\Pi''(\omega,V=0)\approx -\frac{(\beta_L-\beta_R)^2}{2\pi\omega}\frac{4\Gamma_L\Gamma_R}{4\epsilon_d^2+\Gamma^2}=-\frac{(\beta_L-\beta_R)^2}{2\pi\omega}T_d.
\end{align}
In this limit, $\Pi''(\omega)$ has the same structure as for the tunnel junction [see Eq.~\eqref{suscep-final}]. Hence the frequency broadening
\beq
\frac{\delta A}{A}(V=0) \approx -\frac{(\beta_L-\beta_R)^2}{2\pi\omega\kappa}T_d.
\eeq 
Therefore the frequency broadening follows the conductance as a function of the dot gate voltage. The $1/\omega$ term in the frequency series for $\Pi''$ dominates even for a very small asymmetry of the capacitive couplings. Note that, for $\beta_L\approx \beta_R$, the phase shift is still given approximately by  Eq.~\eqref{deltaphi_lorentz} and still has a Lorentzian shape. These equilibrium results (derived for $V=0$) agree with experimental results obtained in Ref.~[\onlinecite{delbecq2011}], where the cavity frequency is the smallest energy scale, and a small asymmetry between the capacitive couplings  is present.


In order to explore how the susceptibility depends on the bias and dot gate voltage, we introduce $\gamma=\displaystyle \left(\beta_L-\lambda\right)/\displaystyle \left(\beta_R-\lambda\right)$ such that
\begin{align}
\frac{\displaystyle \Pi(\omega)}{\displaystyle \left(\beta_L-\lambda\right)\left(\beta_R-\lambda\right)}=&\gamma\Pi_{LL}(\omega)+\frac{\displaystyle \Pi_{RR}(\omega)}{\displaystyle \gamma}  \\
&+\Pi_{LR}(\omega)+\Pi_{RL}(\omega).\nonumber
\end{align}
We remind that for $|\gamma-1|\ll 1$, and $\Gamma_L\sim \Gamma_R$, the out-of-equilibrium relation in Eq.~\eqref{eq:compare} holds. However, no such relation is present between $\delta A/A$  and $G(V)$ at  finite bias.

We plotted both the real and imaginary parts of $\Pi/(\beta_L-\lambda)(\beta_R-\lambda)$ in Fig.~\ref{lowfasym}.
\begin{figure}[t] 
\centering
\includegraphics[width=0.95\linewidth]{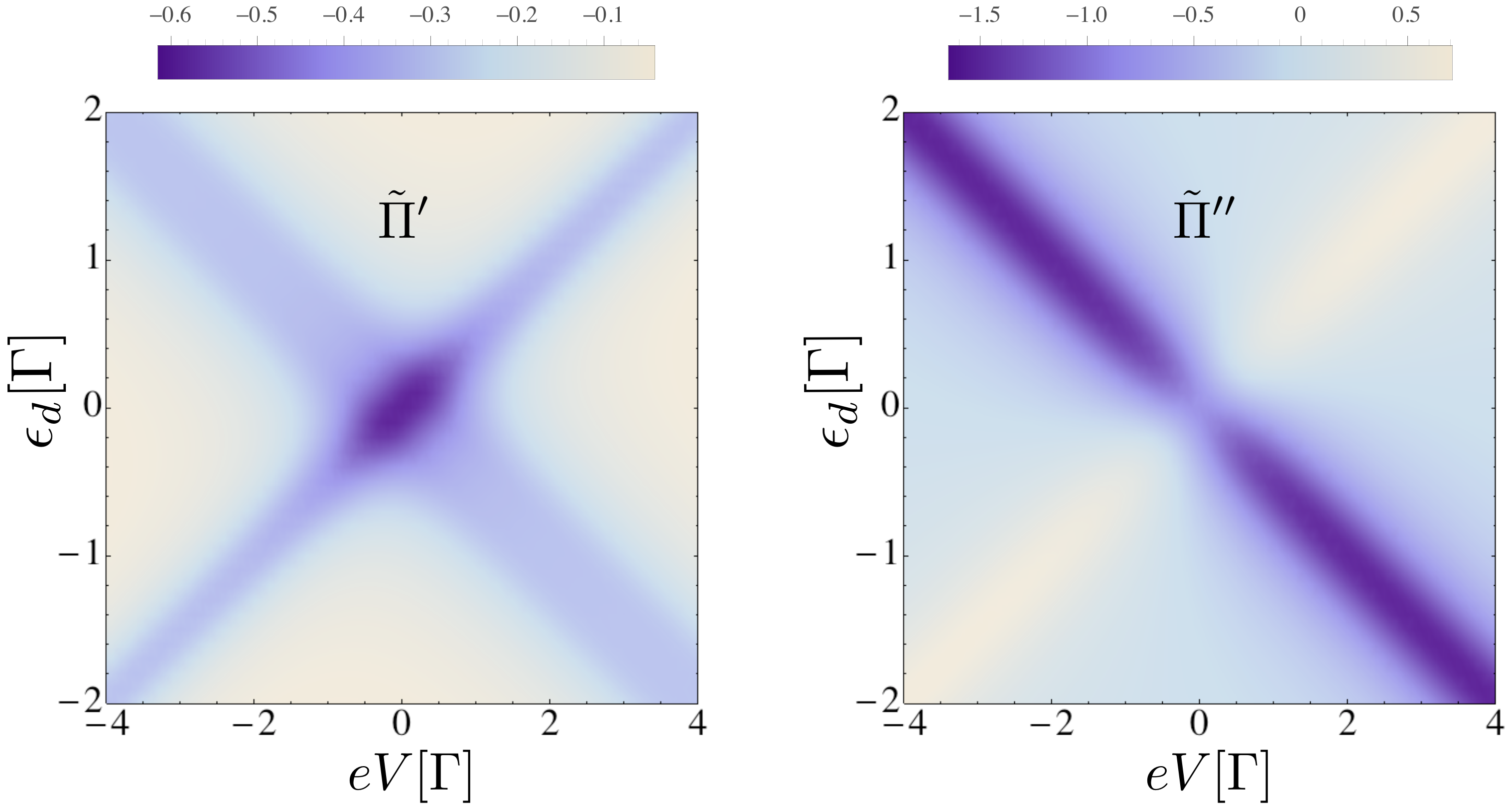}
\caption{$\tilde{\Pi}'(\omega)\equiv\Pi'(\omega)/\gamma\left(\beta_R-\lambda\right)^2$ (left) and $\tilde{\Pi}''(\omega)\equiv\Pi''/\gamma(\beta_R-\lambda)^2$ (right)
as a function of $eV$ and $\epsilon_d$ for  $\Gamma_L=\Gamma_R=0.5$,  $\gamma=0.5$, and $\omega=0.1$. All energies are expressed in terms of $\Gamma=\Gamma_L+\Gamma_R$. 
} 
\label{lowfasym} 
\end{figure}
While $\Pi'$ has the same structure as in Fig.~\ref{lowfsym}, we find that $\Pi''$ has a more complicated non-monotonic behavior.
This is due to the competition between the $\Pi''_{LL}+\Pi''_{RR}$ and $\Pi''_{LR}+\Pi''_{RL}$ terms having opposite signs, which explains, for example, the saddle point in the vicinity of the point $\eps_d=eV=0$. Therefore, in this limit of large asymmetric capacitive coupling, it is no longer possible to relate $\Pi''\propto \delta A/A$ to the differential conductance, even qualitatively.

\subsection{The general case}
Contrary to the previous section where we explored the low frequency regime with cavity frequency being the smallest scale, we now explore the general case with $\omega \sim \Gamma$ making it impossible to expand   $\Pi$ in frequency series.


\subsubsection{Case $\beta_L= \beta_R$}
Analyzing the plots of the real and imaginary parts of $\Pi$ in Fig.~\ref{largefsym},
we observe that $\Pi'$ has now four peaks in the $\eps_d,eV$ plane. Compared to the low frequency case, the two lines of resonance at $eV=\pm 2\eps_d$ are now split into \red{resonances at positions} $eV=\pm 2(\eps_d\pm \omega)$. For such large frequency, the cavity acts like a classical AC voltage. Therefore, it modulates the chemical potential of both leads, which corresponds to the photon assisted tunneling regime. The imaginary part of $\Pi$ in  
Fig.~\ref{largefsym} resembles the one we obtained in the low frequency regime in Fig.~\ref{lowfsym}.

\begin{figure}[t] 
\centering
\includegraphics[width=0.95\linewidth]{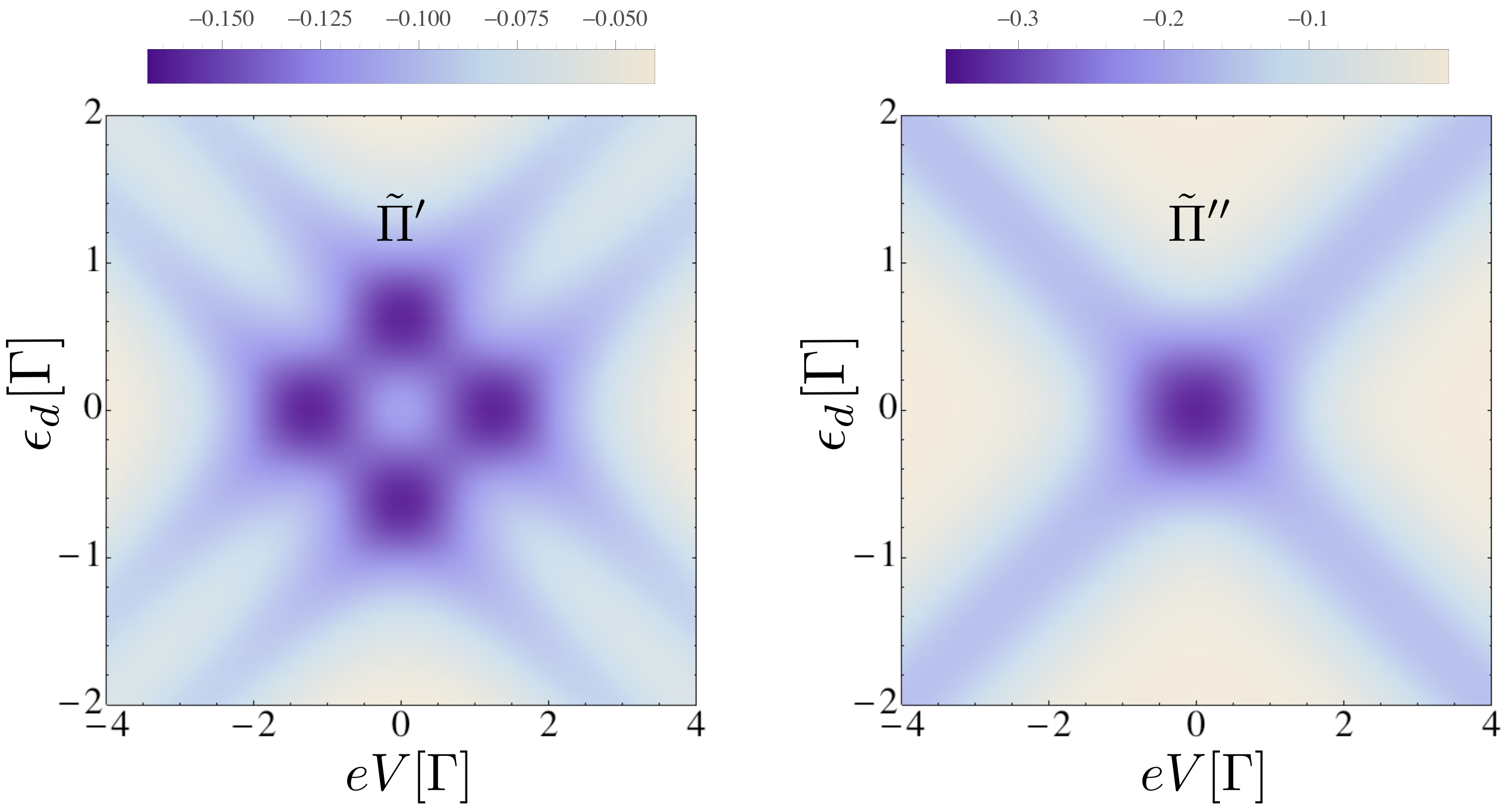}
\caption{Same as Fig.~\ref{lowfsym} for $\omega=0.8$.} 
\label{largefsym} 
\end{figure}
Comparing the resonance lines in both $\Pi'$ and $\Pi''$, we conclude that the Korringa-Shiba relation is violated in this large frequency regime, as expected.

\subsubsection{Case $\beta_L\ne \beta_R$}
Analyzing plots of the real and imaginary part of $\Pi$ for an asymmetry parameter $\gamma=0.5$ in Fig.~\ref{largef_asym},
we observe that $\Pi$ is no longer invariant when $V\to -V $ or $\eps_d\to -\eps_d$.  Let us first focus on the dissipative part $\Pi'$.
In the small $\gamma$ limit, we expect $\Pi/(\beta_L-\lambda)(\beta_R-\lambda)$ to be dominated by $\Pi_{RR}$.
\begin{figure}[h] 
\centering
\includegraphics[width=0.95\linewidth]{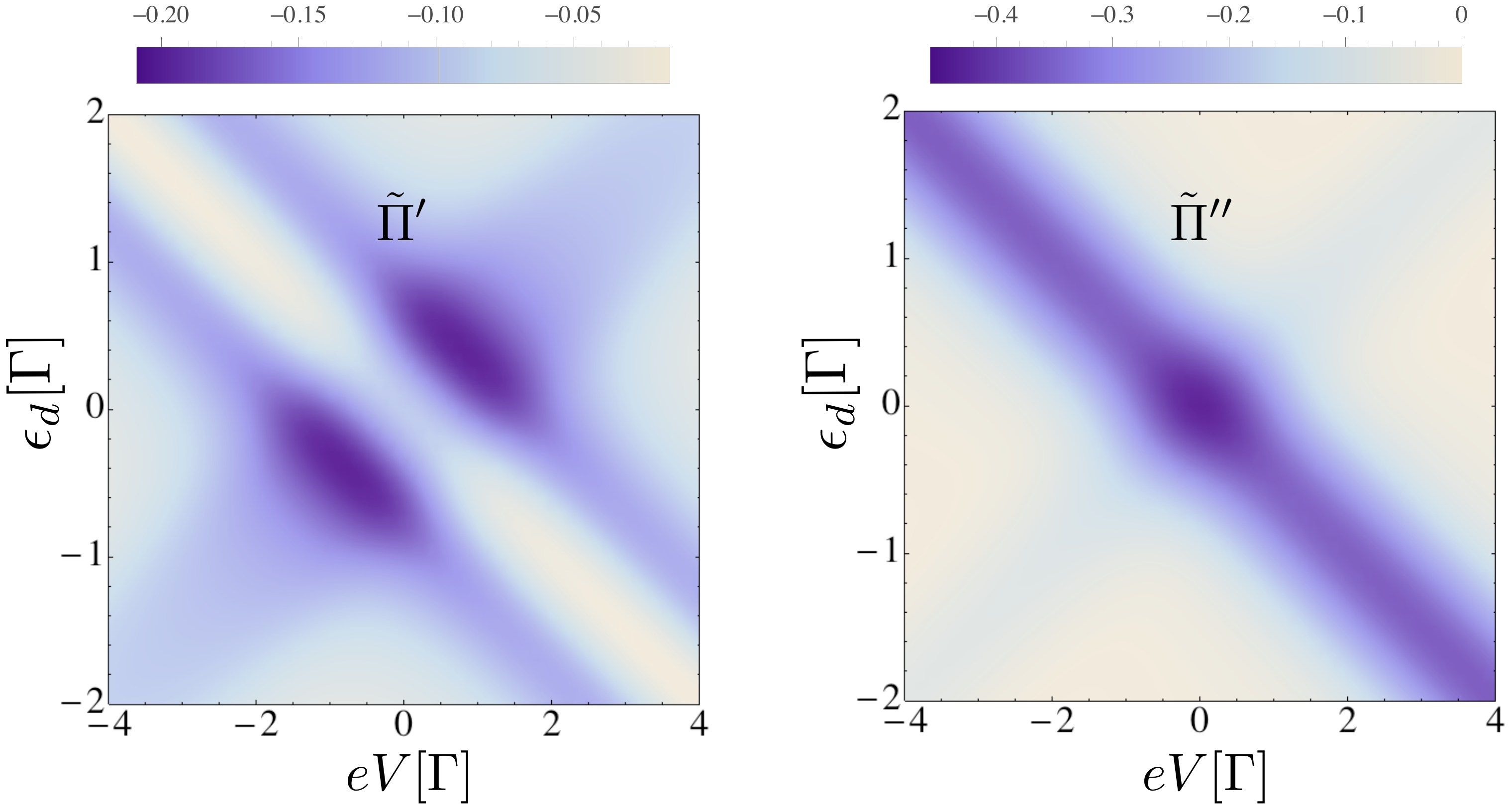}
\caption{Same as Fig. \ref{lowfasym} with  $\omega=0.8$.} 
\label{largef_asym} 
\end{figure}
Since the susceptibility is mainly determined by the right reservoir in this limit, we expect the resonance condition  $-eV/2=\eps_d$ that occurred at small frequency to become
 $-eV/2=\eps_d\pm \omega$ at larger frequency. However, the condition $eV/2=\eps_d$ remains unaltered since the polarizability
is almost blind to the left reservoir. Already for $\gamma=0.5$, we can see \red{the aforemetioned features}  in Fig.~\ref{largef_asym} both for $\Pi'$ and $\Pi''$.

\subsection{A quantum dot in the deep Kondo regime}

As mentioned in Sec.~\ref{sec:susc}, when the quantum conductor is far from any resonance, the susceptibility
and the admittance are related by Eq.~\eqref{relsusceptcond_gen}. This relation holds for a non-equilibrium case and does incorporate the bias $V$. Since $\omega$ is usually fixed by the cavity frequency $\omega_c$, 
$\Pi''(V)$ is proportional to the non-equilibrium differential conductance $Y'(V)$ which is characterized by a Kondo peak around zero bias.  Therefore, we expect the dissipative part of the transmission phase shift to be sensitive to this zero bias anomaly.  However, in the experimental data of Ref.~[\onlinecite{delbecq2011}], the phase shift $\delta \phi (V)\propto \Pi'(V)$ was surprisingly shown to exhibit a behavior  similar to the differential conductance. Extending our analysis of $\Pi(\omega,V)$ to an interacting quantum dot (for example described by the Anderson model) is beyond the scope of the present paper but may help interpreting these experimental results.

\od{

\section{Summary of the results}\label{sec:summary}
We studied a hybrid system consisting of either a tunnel junction or a quantum dot coupled to metallic leads while
the whole system itself is coupled to a single mode of a microwave cavity field. We took into account the capacitive coupling between the cavity and the whole electronic system, \cm{including the leads}.
Using  \cm{an} input-output theory, we found that the optical transmission coefficient is related to the electronic charge susceptibility in the weak electron-photon coupling regime.
Even for non-interacting conductors, many energy scales
\cm{characterize the system}, for example the cavity frequency, the voltage bias, the position and width of a resonance and finally the asymmetry of the capacitive couplings.

For  \cm{a} tunnel junction with \cm{asymmetric} capacitive coupling $\beta_L\neq\beta_R$, the imaginary part of the electonic susceptibility $\Pi(\omega)$ is proportional to the admittance $Y(\omega)$ which is equal to the linear conductance $dI/dV$. 
Similarly, for a quantum dot far from \scrap{a} resonance, therefore characterized by a small conductance and small charge fluctuations, we found that the electronic susceptibility is directly proportional to the quantum dot admittance for all \scrap{the} frequency and voltage. The proportionality constant \cm{is} related to the  \cm{asymmetry} between \cm{the lead and cavity capacitive couplings.} 

The case of symmetric capacitive couplings to the leads, $\beta_L=\beta_R$,  can be absorbed \cm{into} a redefinition of the capacitive coupling to the quantum dot. 
\cm{At zero voltage and small frequency,} we recover the Korringa-Shiba relation that relates the real and imaginary part of the electronic susceptibility\cm{, similarly to the case of}  a quantum dot connected to a single reservoir (\cm{the} so-called quantum RC-circuit). 
\cm{In addition, in the} case of symmetric tunneling amplitudes, $\Gamma_L=\Gamma_R$, we found that the optical phase shift is directly proportional to the differential conductance of the quantum dot  \cm{for arbitrary bias}.
\cm{This strongly symmetric case thus provides a weakly invasive measurement of the non-equilibrium differential conductance, extracted from the cavity field response.}

Nevertheless, the lead capacitive couplings  have no reason to be equal in general since they depend on the particular geometry of the conductor. They are also hardly tunable. 
\cm{Finally, at zero voltage, we find that the imaginary part of the electronic susceptibility is proportional to the equilibrium zero-frequency admittance. Beyond this particular limit however, these two quantities are generally not in correspondance.}

\section{Conclusion and outlook} 
We studied an electronic system consisting of either a tunnel junction or a quantum dot with metallic leads coupled to a single mode of electromagnetic resonator. For weak electron-photon coupling, we related the optical transmission phase shift to the electronic susceptibility.
We have considered and detailed different cases and our results  are summarized in Sec. \ref{sec:summary}.
We demonstrated that  the asymmetry of the capacitive couplings between the electronic reservoirs and the cavity plays actually a crucial role by rendering the cavity electric field  sensitive to charge transfers between two different leads.
We found that the cavity field can be used to probe the low finite frequency admittance $Y_{\alpha\beta}$ (where $\alpha,\beta$ are lead indices) of the electronic system. 
Beyond the low frequency regime, the charge susceptibility and therefore the optical transmission in general offers new insights on the quantum conductor. Since the optical observables are not in direct correspondence with standard transport quantities,  they can be used as a non-invasive probe to better characterize the quantum conductor.

}

Although we  mainly focused in this paper on a non-interacting quantum dot in various regimes, we think that some of our conclusions remain valid for an interacting nanostructure.
As an outlook, it would be interesting to study two out-of-equilibrium quantum dots coupled to the same microwave cavity and use the cavity to both induce and probe the correlated transport through such systems.  Moreover, it would be worth addressing the spin physics using the same approach as developped in this paper, both in the low- and high-frequency regimes.

\section{Acknowledgments}

This work is supported
by a public grant from the ``Laboratoire d'Excellence Physics Atom Light Matter'' (LabEx PALM, reference: ANR-10-LABX-0039) and the French Agence
Nationale de la Recherche through the ANR contract Dymesys.

\appendix

\section{Input-output theory for microwave cavities}\label{appendix_io}
In this section we present details on the derivation of the transmission coefficient of the microwave cavity using input-output theory for the cavity in the presence of the coupling to the quantum dot and metallic leads. We will show that the transmission of the cavity depends on the susceptibility of the electronic system. We consider a single mode cavity with frequency $\omega_c$~\cite{clerk} and, for simplicity, we assume for the moment a single-sided cavity. The total Hamiltonian describing the system reads
\begin{align}
H =\underbrace{H_{el}+H_{ph}+H_{el-c}}_{H_{sys}}+ H_{b} + H_{c-b}\,,
\end{align}
where $H_{sys}$ is defined in the main text in Eq. (4), and 
\begin{align}
H_{b}&=\sum_{q}\hbar\omega_qb^{\dag}_qb_q\,,\\
H_{c-b}&=-i\hbar\sum_{q}\left(f_qa^{\dag}b_q - f^*_qb_q^{\dag}a\right)\,,
\end{align}
are the bath and the cavity-bath Hamiltonians, respectively.  Above, $a$ ($a^\dagger$) is the annihilation (creation) operator for the cavity mode with energy $\omega_c$, $b_{q}$ ($b_{q}^\dagger$) are the annihilation (creation) operators for the bath modes with energy  $\omega_q$, with $q$ labeling  their quantum numbers, and the complex coefficients  $f_{q}$ are coupling parameters between the cavity and the external bath.

The idea of the input-output theory is to find the output photons (or field) in terms of the input ones, as shown schematically in Fig.~\ref{Fig1}. 
 Following Ref.~[\onlinecite{clerk}], we obtain for the cavity equation of motion for a one-side cavity:
\begin{align}
\dot{a} = \frac{\displaystyle i}{\displaystyle \hbar}\left[H_{sys}, a\right] - \frac{\displaystyle \kappa}{\displaystyle 2}a - \sqrt{\kappa}b_{in}\,,
\label{ain}
\end{align}
for the input field, and
\begin{align}
\dot{a} = \frac{\displaystyle i}{\displaystyle \hbar}\left[H_{sys}, a\right] + \frac{\displaystyle \kappa}{\displaystyle 2}a - \sqrt{\kappa}b_{out}\,,
\label{aout}
\end{align}
for the output field, where $\kappa=2\pi\rho|f|^2$ is the cavity decay rate, with $\rho$ the bath density of states and $f$ being the average coupling between the cavity and the bath modes. Subtracting Eq.~\eqref{aout} from Eq.~\eqref{ain} we obtain that
\begin{align}
b_{out}(t) = b_{in}(t) + \sqrt{\kappa}a(t), 
\label{input_output}
\end{align}
a result which holds  for any general cavity Hamiltonian. 

In the following, we will find $b_{out}$ in terms of $b_{in}$ in the presence of the electronic system.
For that, we first evaluate the commutator:
\begin{align}
[H_{sys},a]=-\omega_ca-\lambda n_d-\beta_Ln_L-\beta_Rn_R\,,
\label{eq_motion}
\end{align}
where $n_\alpha$ is the time-dependent electronic particle number ($\alpha=d,L,R$). In order to find the contribution from the electronic system to the equation of motion of the cavity field, we have to utilize the time dependence of one of the electronic particle number operators (let us call it $n_\alpha$ with the coupling constant $\lambda_\alpha$). At time $t$, we can write:

\begin{align}
n_{\alpha H}(t) = U^\dag(t, t_0)n_{\alpha I}(t)U(t, t_0),
\label{nheisenberg}
\end{align}
where
\begin{align}
U(t, t_0) = T_c\exp\left(-i\int_{t_0}^{t}dt'H_{el-c}(t')\right)
\end{align}
is the evolution operator with $T_c$ the time-ordering operator that puts operators with later times to the left of  the ones with earlier times. We can then write Eq.~\eqref{nheisenberg} in the following way
\begin{align}
n_{\alpha H}(t) = n_{\alpha I}(t) + i\lambda_\alpha\int_{t_0}^{t}dt'\Big[(a + a^\dag)n_{\alpha I}(t'), n_{\alpha I}(t)\Big]\,,
\label{noperator}
\end{align}
up to leading order in the coupling constant $\lambda_\alpha$. Thus, the time-evolution of the electronic particle number contains, besides the electronic component, a contribution that arises because of the coupling to the cavity.  Introducing Eq.~\eqref{noperator} into Eq.~\eqref{ain}, assuming that $t_0 \rightarrow -\infty$ and switching to the Fourier space we obtain
\begin{widetext}
\begin{align}
-i\omega a(\omega) = -i\omega_c a(\omega) - \frac{\displaystyle \kappa}{\displaystyle 2}a(\omega) - \sqrt{\kappa}b_{in}(\omega) 
-i\frac{\displaystyle \lambda_\alpha}{\displaystyle \hbar} n_{\alpha I}(\omega) +\nonumber\\
+ \frac{\displaystyle \lambda\alpha^2}{\displaystyle \hbar}\int_{-\infty}^{\infty}dt e^{i\omega t}\int_{-\infty}^{t}dt'\Big[\Big(a(t)e^{-i\omega_c(t' - t)} 
+ a^\dag(t)e^{i\omega_c(t' - t)}\Big)n_{\alpha I}(t'), n_{\alpha I}(t)\Big]\,,
\label{aomega}
\end{align}
\end{widetext}
where $a(t)\approx ae^{-i\omega_ct}$  and $a^\dagger(t)\approx a^\dagger e^{i\omega_ct}$ in zeroth order in $\lambda_\alpha$ (because the expression is already multiplied by $\Lambda\alpha^2$ we can utilize the bare time dependence in this expression). The first term in the above expression describes the free cavity evolution, the second term the leaking into the continuum of modes (the external bath) at rate $\kappa/2$, the third term is the input field coming from the right side, the fourth term correspond to another ``input" contribution to the cavity from the electronic system (a noise term), while the last term leads to both a shift in the cavity frequency as well as to an extra decay channel.  One can now average  over the electronic system, thus neglecting any fluctuation (i.e. feed-back effects).  Moreover, we can neglect  the highly oscillating term $a^\dagger(t)\propto e^{i\omega_ct}$, namely we perform the so called Rotating Wave Approximation (RWA). Under all these assumptions, the last term in Eq.~\eqref{aomega} becomes:
\begin{align}
&\frac{\displaystyle \lambda_\alpha^2}{\displaystyle \hbar}\int_{-\infty}^{\infty}dt e^{i\omega t}\int_{-\infty}^{t}dt'a(t)e^{-i\omega_c(t' - t)}\langle[n_{\alpha I}(t'), n_{\alpha I}(t)]\rangle_0\nonumber\\
&=\od{ -i\int_{-\infty}^{\infty}dt e^{i\omega t}a(t)\Pi(\omega_c) = -ia(\omega)\Pi(\omega_c)},
\label{aomegalastterm2}
\end{align}
where
\begin{align}
\Pi(t' - t) = -i\theta(t' - t)\frac{\displaystyle \lambda_\alpha^2}{\displaystyle \hbar}\langle[n_{\alpha I}(t'), n_{\alpha I}(t)]\rangle_0\,,
\end{align}
is the retarded density-density electronic correlation function utilized in the main text, and $\langle\dots\rangle_0$ means the expectation value of the unperturbed electronic system. \od{Note that for deriving the above expression we assumed that the electronic system is in a stationary state and thus  \cm{time translational invariant}.}

We are now in position to find the cavity field $a(t)$ and the output field $b_{out}(t)$ in terms of the input field $b_{in}(t)$. Introducing Eq.~\eqref{aomegalastterm2} into Eq.~\eqref{aomega} we obtain
\begin{align}
a(\omega) = -\frac{\displaystyle \sqrt{\kappa}b_{in}(\omega)+i(\alpha/\hbar) n_I(\omega)}{\displaystyle -i(\omega - \omega_c) + \kappa/2 \od{+ i\Pi(\omega_c)}}\,.
\label{Res1}
\end{align}

In experiments, one actually encounters a two-sided cavity,
in which case the expression for the cavity equation of motion reads:
\begin{align}
\dot{a} = \frac{\displaystyle i}{\displaystyle \hbar}\left[H_{sys}, a\right] - \left(\frac{\displaystyle \kappa_1}{\displaystyle 2}+\frac{\displaystyle \kappa_2}{\displaystyle 2}\right)a - \sqrt{\kappa_1}b_{in}-\sqrt{\kappa_2}c_{in}\,,
\label{ain2}
\end{align}
so that for the output fields we get:
\begin{align}
b_{out}=\sqrt{\kappa_1}a+b_{in}\\
c_{out}=\sqrt{\kappa_2}a+c_{in}\,.
\end{align}
By following the same reasoning as for the one-sided cavity, we obtain: 
\begin{align}
a(\omega) = -\frac{\displaystyle \sqrt{\kappa_1}b_{in}(\omega) + \sqrt{\kappa_2}c_{in}(\omega)+i(\alpha/\hbar) n_I(\omega)}{\displaystyle -i(\omega - \omega_c) + \kappa_1/2 + \kappa_2/2\od{+ i\Pi(\omega_c)}},
\label{res1twosided}
\end{align}
while if the two mirrors are the same $\kappa_1 = \kappa_2\equiv\kappa$, this becomes
\begin{align}
a(\omega) = -\frac{\displaystyle \sqrt{\kappa}[b_{in}(\omega) + c_{in}(\omega)]+i(\alpha/\hbar) n_I(\omega)}{\displaystyle -i(\omega - \omega_c) + \kappa\od{+ i\Pi(\omega_c)}}\,.
\label{res2twosided}
\end{align}
Assuming again that the input flux is much larger than the electronic contribution, we can write:
\begin{equation}
c_{out}(\omega)=-\tau b_{in}(\omega) + (\ldots) c_{in} (\omega)
\end{equation}
with
\begin{equation}
{\bf \tau}=\frac{\kappa}{{\displaystyle -i(\omega - \omega_c) + \kappa\od{+ i\Pi(\omega_c)}}}\equiv Ae^{i\phi}
\end{equation}
being the transmission of the cavity, which is a complex number, and which depends on the electronic susceptibility $\Pi(\omega)$, as stated in Eq.~1 in the main text.

\section{Scattering matrix approach for a non-interacting quantum conductor}\label{appendix_smatrix}
In this appendix, we express the charge susceptibility of a non-interacting quantum conductor in terms of
integral over the elements of its S matrix.

The electron number  operator in terms of the current operator reads
\begin{align}
n_\alpha(t) = &\frac{\displaystyle i}{\displaystyle 2\pi}\int_{-\infty}^{\infty}d\omega\exp(-i\omega t) \frac{\displaystyle 1}{\displaystyle \omega}\int_{-\infty}^{\infty}dE\nonumber\\
&\sum_{\gamma \gamma '}A_{\gamma \gamma '}(\alpha, E, E + \omega)a^\dag_\gamma(E)a_{\gamma '}(E + \omega),
\label{densityoperator2}
\end{align}
where $A_{\gamma \gamma '}$ have been defined in Eq.~\eqref{A}.
We can now calculate $\Pi_{\alpha\beta}(\omega) \equiv F_{\alpha\beta}(\omega) + X_{\alpha\beta}(\omega)$ with
\begin{align}
F_{\alpha\beta}(t) = &-i\theta(t)\langle n_\alpha(t) n_\beta(0)\rangle,
\label{f1}\\
X_{\alpha\beta}(t) =& i\theta(t)\langle n_\beta(0) n_\alpha(t)\rangle.
\end{align}
Let us first start with $F_{\alpha\beta}(\omega)= -i\int_{0}^{\infty}dte^{i\omega t}\langle n_\alpha(t)n_\beta(0)\rangle$.
Using Eq.~\eqref{densityoperator2} and applying Wick's theorem, we obtain


\begin{align}
&F_{\alpha\beta}(\omega) = \left(\frac{\displaystyle 1}{\displaystyle 2\pi}\right)^2\int_{-\infty}^{\infty}\int_{-\infty}^{\infty}
d\omega_2dE_2 \Bigg\{
\frac{\displaystyle 1}{\displaystyle \omega_2^2}\frac{\displaystyle 1}{\displaystyle \omega + \omega_2 + i\eta}\times\nonumber\\
&\sum_{\gamma_1,\gamma'_1}A_{\gamma_1 \gamma'_1}(\alpha, E_2 + \omega_2, E_2)A_{\gamma'_1 \gamma_1}(\beta, E_2, E_2 + \omega_2)\times\nonumber\\
&\times f_{\gamma_1}(E_2 + \omega_2)\left[1 - f_{\gamma'_1}(E_2)\right]\Bigg\}.
\label{f5}
\end{align}
In the same way we can calculate $X_{\alpha\beta}(\omega)$:
\begin{align}
&X_{\alpha\beta}(\omega) = -\left(\frac{\displaystyle 1}{\displaystyle 2\pi}\right)^2\int_{-\infty}^{\infty}\int_{-\infty}^{\infty}
d\omega_2dE_2 \Bigg\{
\frac{\displaystyle 1}{\displaystyle \omega_2^2}\frac{\displaystyle 1}{\displaystyle \omega + \omega_2 + i\eta}\nonumber\\
\!\!&\times\sum_{\gamma_1,\gamma'_1}A_{\gamma_1 \gamma'_1}(\alpha, E_2 + \omega_2, E_2)A_{\gamma'_1 \gamma_1}(\beta, E_2, E_2 + \omega_2)\times\nonumber\\
&\times f_{\gamma'_1}(E_2)\left[1 - f_{\gamma_1}(E_2 + \omega_2)\right]\Bigg\}.
\label{x2}
\end{align}
And finally $\Pi_{\alpha\beta}(\omega)$ reads

\begin{align}
&\Pi_{\alpha\beta}(\omega) = \left(\frac{\displaystyle 1}{\displaystyle 2\pi}\right)^2\int_{-\infty}^{\infty}\int_{-\infty}^{\infty}d\omega_2dE_2
\left\{
\frac{\displaystyle 1}{\displaystyle \omega_2^2}\frac{\displaystyle 1}{\displaystyle \omega + \omega_2 + i\eta}\times\right.\nonumber\\
&\left.\sum_{\gamma_1,\gamma'_1}F^{\alpha\beta}_{\gamma_1\gamma'_1}(E_2, \omega_2)\left[f_{\gamma_1}(E_2 + \omega_2) - f_{\gamma'_1}(E_2)\right]\right\}.
\label{corrfuncresult2b}
\end{align}
where we introduced
\begin{align}
F^{\alpha\beta}_{\gamma\gamma'} = A_{\gamma\gamma'}(\alpha, E + \omega, E)A_{\gamma'\gamma}(\beta, E, E + \omega).
\end{align}

 At zero temperature, we can further simplify Eq.~\eqref{corrfuncresult2b} and using
%
the identity
\begin{align}
\frac{\displaystyle 1}{\displaystyle \omega + \omega_2 + i\eta} = \mathcal{P}\left(\frac{1}{\omega + \omega_2}\right) - i\pi\delta(\omega + \omega_2)\,,
\label{principalvalue}
\end{align}
we obtain Eq.~\eqref{dalphabeta} and Eq.~\eqref{palphabeta} of the main text.

\section{Calculation of the quantum dot susceptibility}\label{appendix_susc}

\subsection{Calculation of the real part of the susceptibility}
In this section we give technical details on how to compute the real part of the charge susceptibility. Let us compute one of the integrals
\begin{align}
L(y) = 2i\mathcal{P}\int_{-\infty}^{\infty}\frac{\displaystyle dx}{\displaystyle x^2(x+y)(x^2+4)}f(x),
\label{int1}
\end{align}
where
\begin{align}
f(x)=\arctan(a+x)-\arctan(a-x).
\end{align}

Using the decomposition
\begin{align}
\arctan(\alpha + x) = \frac{\displaystyle 1}{\displaystyle 2 i} \left( \log[1 + i (\alpha+x)] - \log[1 - i (\alpha+x)] \right),
\end{align}
the function $f(x)$ can be given the following alternative form
\begin{align}
f(x)=&\frac{\displaystyle 1}{\displaystyle 2i}\left[\log\left(1+\frac{\displaystyle ix}{\displaystyle 1+ia}\right)+\log\left(1+\frac{\displaystyle ix}{\displaystyle 1-ia}\right) \right.\nonumber\\
&-\left.\log\left(1-\frac{\displaystyle ix}{\displaystyle 1+ia}\right)-\log\left(1-\frac{\displaystyle ix}{\displaystyle 1-ia}\right)\right].
\end{align}
Therefore, $L(y)$ can be written as a sum over four integrals. Let us detail the calculation of one of them.
We consider the following principal part integral
\begin{align}
L_1(y) = &\mathcal{P}\int_{-\infty}^{\infty}\frac{\displaystyle dx}{\displaystyle x^2(x+y)(x+2i)(x-2i)}\nonumber\\
&\times\log\left(1+\frac{\displaystyle ix}{\displaystyle 1+ia}\right).
\label{int2}
\end{align}
$L_1(y)$ can be computed with standard complex plane integration techniques. The integrand has five singularities in the complex plane: poles at $z=0$, $z=-y$, $z=-2i$, $z=2i$ and a branch point
at $z = i-a$. A semi-circle is added around $z=0$ and $z=-y$. The contour is closed in the lower half-plane, avoiding both the pole at $z= 2i$ and the branch cut [Fig.~\ref{poles}]. 
\begin{figure}[b] 
\centering
\includegraphics[width=0.8\linewidth]{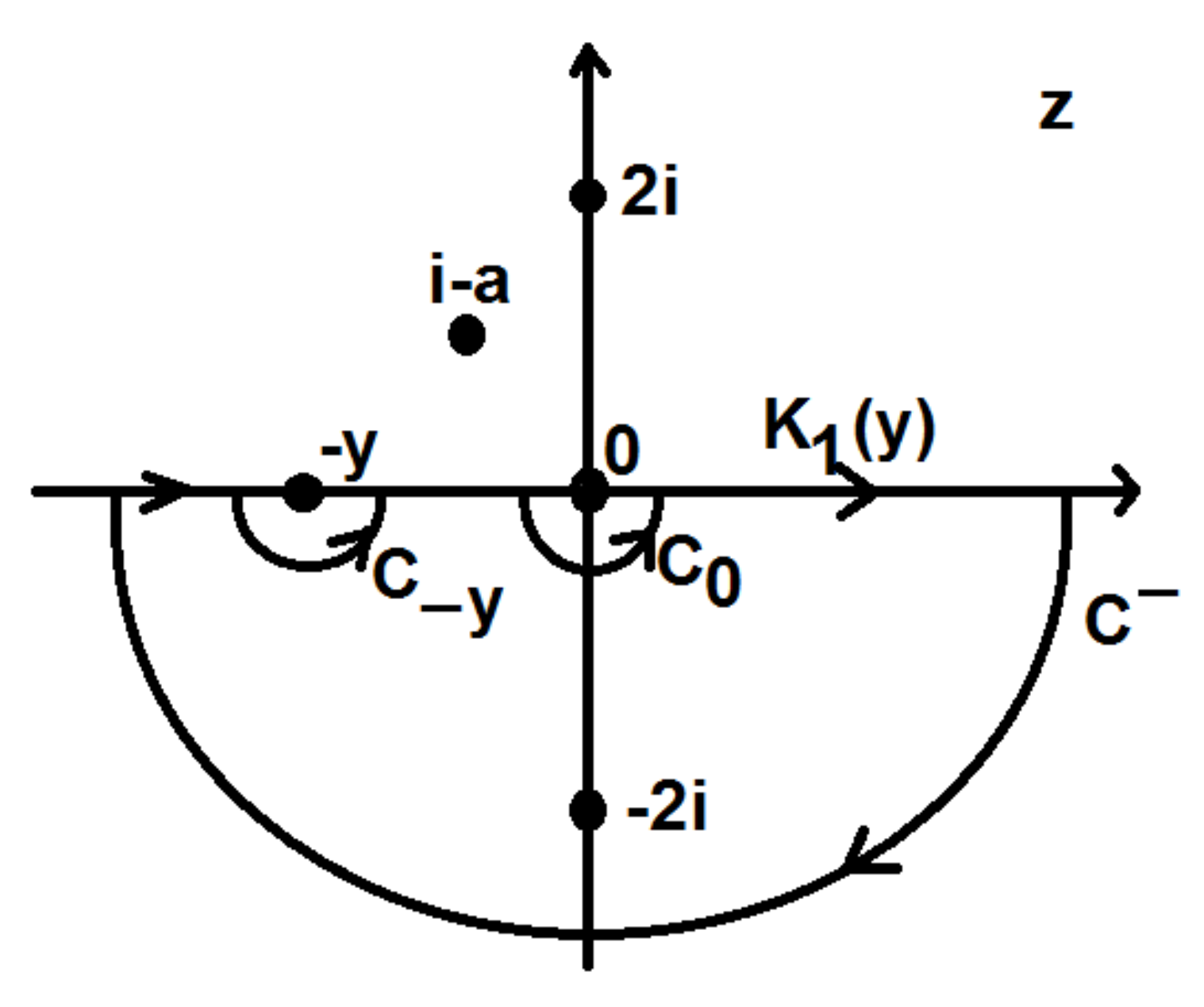}
\caption{Complex plane $z$ with poles of the integrand, Eq.(\ref{int2}).}
\label{poles} 
\end{figure}

Using Cauchy's residue theorem, we obtain eventually: 
\begin{align}
-2\pi i \text{Res}[-2i]=\int_{C^-}=L_1(y)+\int_{C_{-y}}+\int_{C_{0}}+\int_{C_R}
\end{align}
where
\begin{align}
L_1(y)=-\pi i \left(2\text{Res}[-2i]+\text{Res}[-y]+\text{Res}^{(2)}[0]\right),
\end{align}
and
\begin{align}
\text{Res}[-y]=
\frac{\displaystyle 1}{\displaystyle y^2(y^2+4)}\log\left(1-\frac{\displaystyle iy}{\displaystyle 1+ia}\right),
\end{align}
\begin{align}
\text{Res}[-2i]=\frac{\displaystyle 1}{\displaystyle 16i(y-2i)}\log\left(1+\frac{\displaystyle 2}{\displaystyle 1+ia}\right),
\end{align}
\begin{align}
\text{Res}^{(2)}[0]=\frac{\displaystyle i}{\displaystyle 4y(1+ia)}\,.
\end{align}
Above,  Res$[a]\equiv {\rm Res}(f,a)=\lim_{x\rightarrow a}(z-a)f(z)$ and ${\rm Res}^{(2)}[a]\equiv {\rm Res}^{(2)}(f,a)=\lim_{x\rightarrow a}d/dz[(z-a)^2f(z)]$ are the residues of the function $f(z)$ at the pole  $a$ of the first and second order, respectively.  Summing everything, we finally obtain
\begin{widetext}
\begin{align}
L_1(y)=-\pi i\left[\frac{\displaystyle 1}{\displaystyle 8i(y-2i)}\log\left(1+\frac{\displaystyle 2}{\displaystyle 1+ia}\right)+\frac{\displaystyle 1}{\displaystyle y^2(y^2+4)}\log\left(1-\frac{\displaystyle iy}{\displaystyle 1+ia}\right)+\frac{\displaystyle i}{\displaystyle 4y(1+ia)}\right].
\end{align}
The other terms appearing in $L(y)$ can be calculated using a similar approach.

\subsection{General expression for the quantum dot susceptibility}
Gathering all terms, the real and imaginary parts of the susceptibility read

\begin{align}
&\Pi'_{LL}(\omega)=-\frac{\displaystyle 4\Gamma_L^2}{\displaystyle \pi\Gamma^3y\left(y^2+4\right)}\left[\arctan\left(a+y\right)-\arctan\left(a-y\right)\right]\nonumber\\
\!\!&+\frac{\displaystyle \Gamma_L}{\displaystyle \pi\Gamma^2\left(y^2+4\right)}\left(1+\frac{\displaystyle 4\Gamma_R}{\displaystyle y^2\Gamma}\right)\left[-2\log\left(1+a^2\right)+\log\left[1+\left(a+y\right)^2\right]+\log\left[1+\left(a-y\right)^2\right]\right],
\label{pllrealgen}
\end{align}

\begin{align}
&\Pi''_{LL}(\omega)=-\frac{\displaystyle 2\Gamma_L}{\displaystyle \pi\Gamma^2\left(y^2+4\right)}\left(1+\frac{\displaystyle 4\Gamma_R}{\displaystyle y^2\Gamma}\right)\left[\arctan\left(a+y\right)-\arctan\left(a-y\right)\right]\nonumber\\
\!\!&-\frac{\displaystyle 2\Gamma_L^2}{\displaystyle \pi\Gamma^3y\left(y^2+4\right)}\left[-2\log\left(1+a^2\right)+\log\left[1+\left(a+y\right)^2\right]+\log\left[1+\left(a-y\right)^2\right]\right],
\label{pllimgen}
\end{align}

\begin{align}
&\Pi'_{RR}(\omega)=-\frac{\displaystyle 4\Gamma_R^2}{\displaystyle \pi\Gamma^3y\left(y^2+4\right)}\left[\arctan\left(b+y\right)-\arctan\left(b-y\right)\right]\nonumber\\
\!\!&+\frac{\displaystyle \Gamma_R}{\displaystyle \pi\Gamma^2\left(y^2+4\right)}\left(1+\frac{\displaystyle 4\Gamma_L}{\displaystyle y^2\Gamma}\right)\left[-2\log\left(1+b^2\right)+\log\left[1+\left(b+y\right)^2\right]+\log\left[1+\left(b-y\right)^2\right]\right],
\end{align}

\begin{align}
&\Pi''_{RR}(\omega)=-\frac{\displaystyle 2\Gamma_R}{\displaystyle \pi\Gamma^2\left(y^2+4\right)}\left(1+\frac{\displaystyle 4\Gamma_L}{\displaystyle y^2\Gamma}\right)\left[\arctan\left(b+y\right)-\arctan\left(b-y\right)\right]\nonumber\\
\!\!&-\frac{\displaystyle 2\Gamma_R^2}{\displaystyle \pi\Gamma^3y\left(y^2+4\right)}\left[-2\log\left(1+b^2\right)+\log\left[1+\left(b+y\right)^2\right]+\log\left[1+\left(b-y\right)^2\right]\right],
\end{align}

\begin{align}
&\Pi'_{LR}(\omega)+\Pi'_{RL}(\omega)=-\frac{\displaystyle 4\Gamma_L\Gamma_R}{\displaystyle \pi\Gamma^3y\left(y^2+4\right)}\left[\arctan\left(a+y\right)-\arctan\left(a-y\right)\right]\nonumber\\
\!\!&-\frac{\displaystyle 4\Gamma_L\Gamma_R}{\displaystyle \pi\Gamma^3y^2\left(y^2+4\right)}\left[-2\log\left(1+a^2\right)+\log\left[1+\left(a+y\right)^2\right]+\log\left[1+\left(a-y\right)^2\right]\right]\nonumber\\
\!\!&-\frac{\displaystyle 4\Gamma_L\Gamma_R}{\displaystyle \pi\Gamma^3y\left(y^2+4\right)}\left[\arctan\left(b+y\right)-\arctan\left(b-y\right)\right]\nonumber\\
\!\!&-\frac{\displaystyle 4\Gamma_L\Gamma_R}{\displaystyle \pi\Gamma^3y^2\left(y^2+4\right)}\left[-2\log\left(1+b^2\right)+\log\left[1+\left(b+y\right)^2\right]+\log\left[1+\left(b-y\right)^2\right]\right],
\label{plrrlrealgen}
\end{align}

\begin{align}
&\Pi''_{LR}(\omega)+\Pi''_{RL}(\omega)=\frac{\displaystyle 8\Gamma_L\Gamma_R}{\displaystyle \pi\Gamma^3y^2\left(y^2+4\right)}\left[\arctan\left(a+y\right)-\arctan\left(a-y\right)\right]\nonumber\\
\!\!&-\frac{\displaystyle 2\Gamma_L\Gamma_R}{\displaystyle \pi\Gamma^3y\left(y^2+4\right)}\left[-2\log\left(1+a^2\right)+\log\left[1+\left(a+y\right)^2\right]+\log\left[1+\left(a-y\right)^2\right]\right]\nonumber\\
\!\!&+\frac{\displaystyle 8\Gamma_L\Gamma_R}{\displaystyle \pi\Gamma^3y^2\left(y^2+4\right)}\left[\arctan\left(b+y\right)-\arctan\left(b-y\right)\right]\nonumber\\
\!\!&-\frac{\displaystyle 2\Gamma_L\Gamma_R}{\displaystyle \pi\Gamma^3y\left(y^2+4\right)}\left[-2\log\left(1+b^2\right)+\log\left[1+\left(b+y\right)^2\right]+\log\left[1+\left(b-y\right)^2\right]\right].
\label{plrrlimgen}
\end{align}

Using 
\begin{align}
\Pi(\omega) = \left(\beta_L - \lambda\right)^2\Pi_{LL}(\omega) + \left(\beta_R - \lambda\right)^2\Pi_{RR}(\omega) +\left(\beta_L - \lambda\right)\left(\beta_R - \lambda\right)\left[\Pi_{LR}(\omega) + \Pi_{RL}(\omega)\right],
\end{align}
 we obtain the final result for $\Pi(\omega)$:

\begin{align}\label{eq:pi_r}
&\Pi'(\omega)=-\frac{\displaystyle 4\Gamma_L}{\displaystyle \pi\Gamma^3y\left(y^2+4\right)}\left[\arctan\left(a+y\right)-\arctan\left(a-y\right)\right]\left(\left(\beta_L - \lambda\right)^2\Gamma_L+\left(\beta_L - \lambda\right)\left(\beta_R - \lambda\right)\Gamma_R\right)\nonumber\\
\!\!&+\frac{\displaystyle \Gamma_L}{\displaystyle \pi\Gamma^2\left(y^2+4\right)}\left[-2\log\left(1+a^2\right)+\log\left[1+\left(a+y\right)^2\right]+\log\left[1+\left(a-y\right)^2\right]\right]\nonumber\\
\!\!&\times
\Bigg(\left(\beta_L - \lambda\right)^2\left(1+\frac{\displaystyle 4\Gamma_R}{\displaystyle y^2\Gamma}\right)-\left(\beta_L - \lambda\right)\left(\beta_R - \lambda\right)\frac{\displaystyle 4\Gamma_R}{\displaystyle y^2\Gamma}\Bigg)\nonumber\\
\!\!&-\frac{\displaystyle 4\Gamma_R}{\displaystyle \pi\Gamma^3y\left(y^2+4\right)}\left[\arctan\left(b+y\right)-\arctan\left(b-y\right)\right]\left(\left(\beta_L - \lambda\right)^2\Gamma_R+\left(\beta_L - \lambda\right)\left(\beta_R - \lambda\right)\Gamma_L\right)\nonumber\\
\!\!&+\frac{\displaystyle \Gamma_R}{\displaystyle \pi\Gamma^2\left(y^2+4\right)}\left[-2\log\left(1+b^2\right)+\log\left[1+\left(b+y\right)^2\right]+\log\left[1+\left(b-y\right)^2\right]\right]\nonumber\\
\!\!&\times
\Bigg(\left(\beta_L - \lambda\right)^2\left(1+\frac{\displaystyle 4\Gamma_L}{\displaystyle y^2\Gamma}\right)-\left(\beta_L - \lambda\right)\left(\beta_R - \lambda\right)\frac{\displaystyle 4\Gamma_L}{\displaystyle y^2\Gamma}\Bigg).
\end{align}
and the imaginary part of the total charge susceptibility reads
\begin{align}\label{eq:pi_i}
&\Pi''(\omega)=-\frac{\displaystyle 2\Gamma_L}{\displaystyle \pi\Gamma^2\left(y^2+4\right)}\left[\arctan\left(a+y\right)-\arctan\left(a-y\right)\right]
\Bigg(\left(\beta_L - \lambda\right)^2\left(1+\frac{\displaystyle 4\Gamma_R}{\displaystyle y^2\Gamma}\right)-\left(\beta_L - \lambda\right)\left(\beta_R - \lambda\right)\frac{\displaystyle 4\Gamma_R}{\displaystyle y^2\Gamma}\Bigg)\nonumber\\
\!\!&-\frac{\displaystyle 2\Gamma_L}{\displaystyle \pi\Gamma^3y\left(y^2+4\right)}\left[-2\log\left(1+a^2\right)+\log\left[1+\left(a+y\right)^2\right]+\log\left[1+\left(a-y\right)^2\right]\right]
\left(\left(\beta_L - \lambda\right)^2\Gamma_L+\left(\beta_L - \lambda\right)\left(\beta_R - \lambda\right)\Gamma_R\right)\nonumber\\
\!\!&-\frac{\displaystyle 2\Gamma_R}{\displaystyle \pi\Gamma^2\left(y^2+4\right)}\left[\arctan\left(b+y\right)-\arctan\left(b-y\right)\right]
\Bigg(\left(\beta_L - \lambda\right)^2\left(1+\frac{\displaystyle 4\Gamma_L}{\displaystyle y^2\Gamma}\right)-\left(\beta_L - \lambda\right)\left(\beta_R - \lambda\right)\frac{\displaystyle 4\Gamma_L}{\displaystyle y^2\Gamma}\Bigg)\nonumber\\
\!\!&-\frac{\displaystyle 2\Gamma_R}{\displaystyle \pi\Gamma^3y\left(y^2+4\right)}\left[-2\log\left(1+b^2\right)+\log\left[1+\left(b+y\right)^2\right]+\log\left[1+\left(b-y\right)^2\right]\right]
\left(\left(\beta_L - \lambda\right)^2\Gamma_R+\left(\beta_L - \lambda\right)\left(\beta_R - \lambda\right)\Gamma_L\right).
\end{align}
\end{widetext}
where we remind that we introduced the dimensionless variables
$
a=\frac{\displaystyle eV-2\epsilon_d}{\displaystyle \Gamma},~~
b=\frac{\displaystyle 2\epsilon_d+eV}{\displaystyle \Gamma},~~{\rm and}~~
y=\frac{\displaystyle 2\omega}{\displaystyle \Gamma}$.

\bibliography{bibliography}

\end{document}